\documentclass{aa}
\usepackage{graphics,natbib}
\newcommand{\um}{\,$\mu$m}
\newcommand{\cyga}{Cygnus~A}

\begin{document}

\title{New radiative transfer models for obscuring tori in active galaxies}

\author{I.M. van Bemmel\inst{1} \and C.P. Dullemond\inst{2}}

\institute{Kapteyn Astronomical Institute, P.O.Box 800, NL--9700 AV Groningen
  \and
  Max-Planck Institut f\"ur Astronomie, P.O.Box 1317, D--85741 Garching}

\offprints{I.M. van Bemmel (bemmel@stsci.edu)}

\date{Received date; Accepted date}
\authorrunning{van Bemmel \& Dullemond}

\abstract{Two-dimensional radiative transfer is employed to obtain the 
broad-band infrared spectrum of active galaxies. In the models we 
vary the geometry and size of the obscuring medium, the surface 
density, the opacity and the grain size distribution. Resulting
spectral energy distributions are constructed for different 
orientations of the toroid. Colour-colour comparisons with 
observational data are consistent with previous observations that
the emission longward of 60\um\ is produced by star-formation and
unrelated to the presence of an obscuring torus.
We also find that the toroid cannot be flat, but is rather conical 
or flaring. The density is most likely constant with radius, and 
the size is relatively large with an inner radius around 10\,pc. 
A direct comparison with radio galaxy \cyga\ yields a best fit for 
a conical disk with constant surface density, and a size from 10 
to 30\,pc, assuming the far-infrared emission is due to star-formation 
in the host galaxy.
\keywords{galaxies:active, galaxies:Seyfert, galaxies:starburst, 
galaxies individual: \cyga, 
infrared:galaxies, infrared:ISM, quasars:general, ISM:dust}}

\maketitle

\section{Introduction}
Unified models for active galaxies
postulate the existence of a circumnuclear dust torus. This 
torus obscures the central regions 
from view at certain solid angles, at wavelengths
where the dust is optically thick. Various authors
have studied the behaviour of a dusty torus in the presence of a
strong radiation field. Models have been constructed for the
absorption of the optical to FUV emission by the
dust, and the re-emission of this energy in the infrared
\citep{pk92,gra94,efst95}. These models
have been used to interpret IRAS observations of active galaxies.
Their main result is that the spectrum from a dusty disk is
much broader than a single grey body. However, observed SEDs are
often still broader than the models,  and show different
behaviour of the 10\um\ silicate feature. 

As discussed in previous work \citep{ivb01}, it is very likely that there
are two dust components in active galaxies. One component is a
large scale dust distribution, heated by stars. The other is the
obscuring torus. In narrow-line radio galaxies and quasars it
is difficult to disentangle the two components, since their
broad-band spectra are not well known, due to their distance
and the sensitivity of previous infrared instruments.
Previous model studies of obscuring tori were only concerned 
with another subset of the active galaxies, namely Seyfert
galaxies, which happen to be nearby and therefore bright and
well observed infrared sources. Moreover, these studies focused 
on reproducing the general width of the SED. Furthermore, the 
emphasis was on fitting a
few selected objects, rather than a complete sample. 
In previous studies, the model geometry has been fixed, and
the dependence of the resulting SED on the geometry and
other disk parameters is not well understood.
Finally, the models assume standard Galactic dust, 
which might not be appropriate for luminous active galaxies. 
Several studies \citep[e.g.][]{laor93,maio01_2,maio01_1} have
shown that the lack of a 10\um\ silicate emission feature and
shallow 2200\,\AA\ absorption can point to dust that is dominated 
by relatively large grains, while Galactic dust
is dominated by small grains. 

\begin{table*}[!Ht]
\begin{center}
\begin{tabular}{|l|c|c|c|}
\hline 
Parameter & Symbol & Values & Reference\\
	&	&	& model \\
\hline
Equatorial optical depth at 20\um & $\tau_{20}$ & 5, 10, 20, 50 & 10 \\
Scale height exponent  & $\alpha$ & 0, 1, 9/7 & 1 \\
Surface density exponent & $\beta$ & 0, --1, --2 & 0 \\
Inner radius of the disk (pc) & $r_0$ & 3, 30 & 3, 30 \\
Outer radius of the disk (pc) & $r_{\rm out}$ & 30, 300 & 30, 300 \\
Aspect angle (degrees)& $\theta$ & 5, 30, 50, 60, 70, 90 & all \\
Grain size exponent & $q$ & --2, --3.5 & --2 \\
Scale height at inner radius (pc) & $h_0$ & $=r_0$ & 3 \\
Spectral index of input spectrum & $j$ & --2 & --2 \\
100--400\,eV luminosity & $L_{X}$ & $10^{12}$\,L$_{\odot}$ & $10^{12}$\,L$_{\odot}$ \\
\hline
\end{tabular}
\caption{\label{7_model_params}
The different values for the model parameters, and their
value in the reference model. The inclination $\theta$ is defined 
with respect to the symmetry axis of the torus.}
\end{center}
\end{table*}

More recent models treat the torus as a clumpy medium, which
produces a broader infrared spectrum, more consistent with
the observations \citep{nenk02}. However, these
authors do not vary the geometry of the torus, nor the distribution
of the mass. Although it is likely that the torus does not have a smooth
density distribution, it is not yet understood how other
parameters influence the width of the resulting broad-band
spectrum. Moreover, they assume that all infrared emission 
observed in active galaxies arises in the torus.

Here we will re-examine the infrared SEDs of active galaxies from
a theoretical point of view. We employ a new and very versatile
radiative transfer code {\sc radical} \citep{dul00},
and examine a wide parameter space. Specifically, we study the 
dependence of the emission on observing angle, optical depth, disk
geometry and size, dust mass distribution and dust grain size
distribution. The initial goal will be to constrain the
parameters of the torus, by comparing the colours to observations,
and by fitting the infrared broad-band spectrum of \cyga. As a
first order approximation we will assume the torus has a smooth
distribution of dust. 
Eventually, these new models can be used to disentangle the two dust 
components in active galaxies, which will allow us to study 
star-formation in active galaxies, and the parameters of the 
obscuring torus in more detail than before.

\section{Model setup}
In our disk model a geometry for the dust, an input spectrum and
dust opacities need to be specified. The numerical method is 
described below in
Sec.~\ref{7_radical}. We have constructed a grid of models, 
built by varying different parameters. This results in
a set of over 150 model SEDs, in which all variables
are cross-correlated. Below we describe the different parameters 
of the models (see also Table~\ref{7_model_params}).

\subsection{Disk geometry and density distribution}
The density of the disk in angular direction is a Gaussian,
with a width of $h(r)$. The mass density for any point in 
the disk is then given by:
\begin{displaymath}
\rho({\rm r}, \theta) = \frac{\Sigma({\rm r})}{h({\rm r}) \sqrt{\pi}}
\cdot \exp{\left[-\left(\frac{z}{h({\rm r})}\right)^2\right]}
\end{displaymath}
with $z = r*\tan(\theta)$ and $\Sigma(r)$ the surface density.
Due to the setup of the model, the surface where $\tau_{20}$ 
equals unity does not coincide with the scale height of the
disk. For higher opacities this surface lies at higher $z$.
Therefore the actual opening angle of the torus is smaller 
for higher opacities, although the width of the Gaussian is 
kept constant.

The geometry of the disk is thus described by four parameters: the scale 
height $h(r)$, the surface density $\Sigma(r)$, the inner 
radius $r_0$ and the outer radius $r_{\rm out}$. 
Each of these parameters is varied. For the scale height we assume:
\begin{displaymath}
h(r) = h_0 \left(\frac{r}{r_0}\right)^{\alpha} 
\end{displaymath}
and the surface density is defined as:
\begin{displaymath}
\Sigma(r) = \Sigma_0 \left(\frac{r}{r_0}\right)^{\beta} 
\end{displaymath}
with $r_0$ the inner radius of the disk.

For $h_0$ we assume it equals the inner radius, i.e. $h(r_0)=h_0=r_0$,
so the disk is thicker 
if it is further away from the emission source. This implies that
the opening angles are comparable for similar values of
$\tau_{20}$. The resulting opening angles are consistent
with the size of the broad-line region that has to be shielded in
a more edge-on orientation, and with the observed opening angles
of ionization cones in active galaxies. 

\begin{figure*}[!Ht]
\resizebox{8.7cm}{!}{\includegraphics{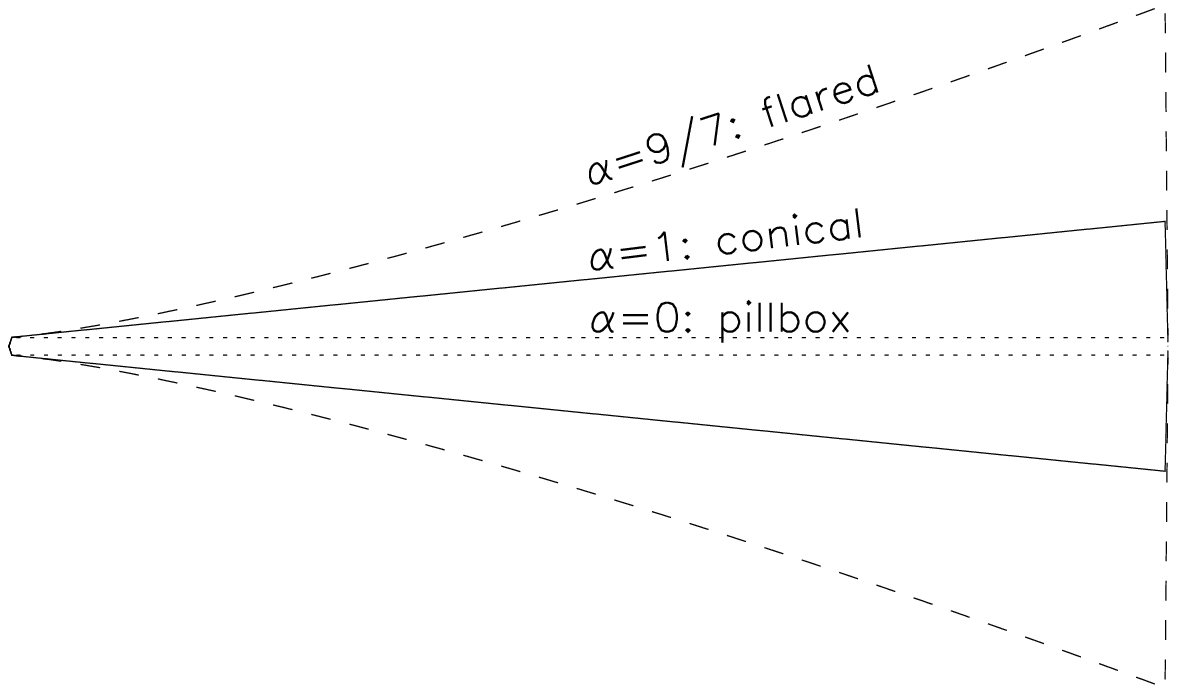}}
\resizebox{8.7cm}{!}{\includegraphics{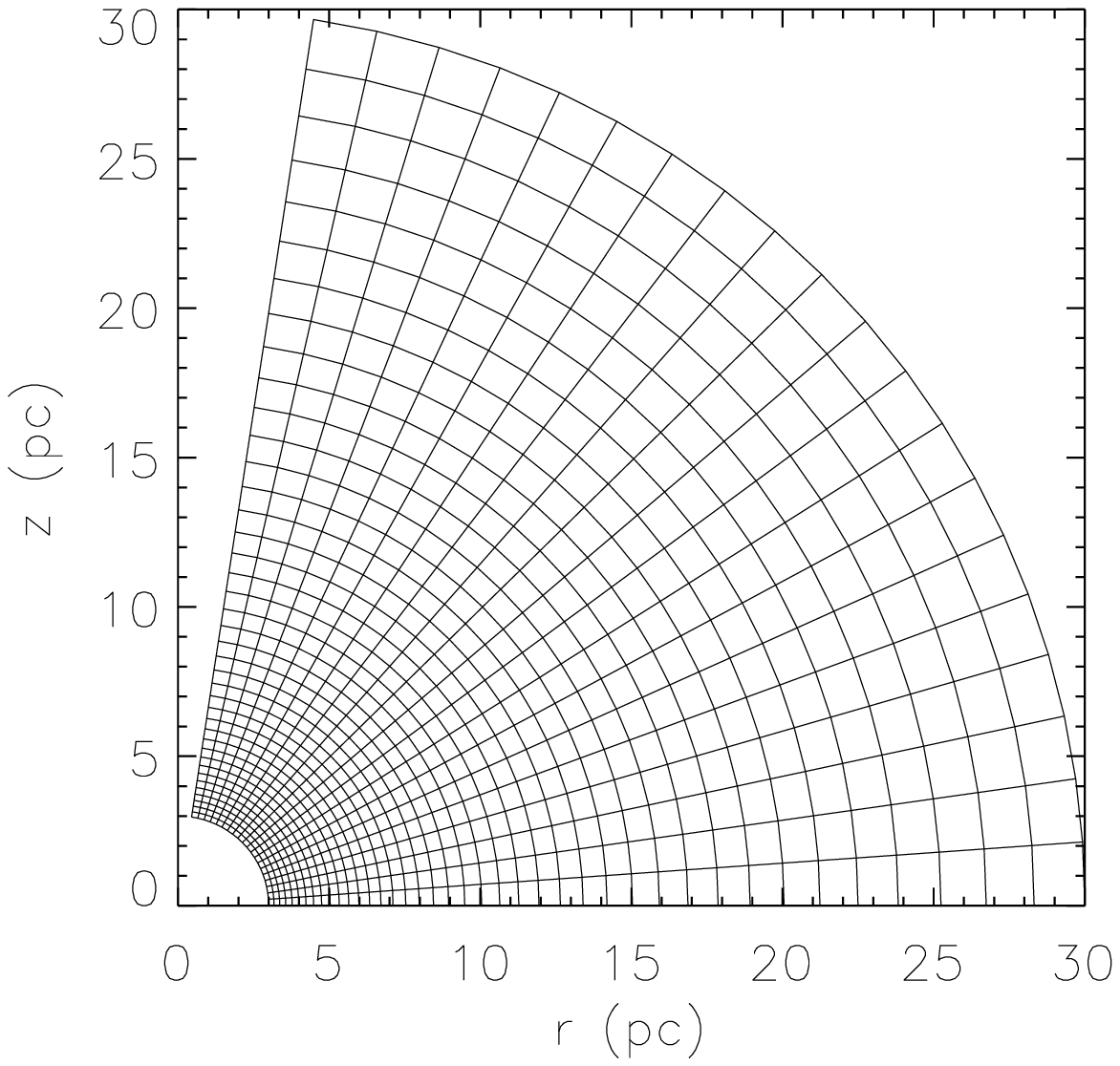}}
\caption{\label{7_geom}
{\it Left panel}: the three different geometries of the dust disk. Dotted
line is the pillbox model ($\alpha=0$), solid line the
conical disk ($\alpha=1$), and the dashed line is the
flared disk ($\alpha=9/7$).
{\it Right panel}: an illustration of the grid setup in Cartesian coordinates. The
grid is mirrored in the $r$-axis and rotated on the $z$-axis to
achieve full spatial coverage of the dust disk. Refinement of the
grid cells at small radii is not shown.}
\end{figure*}

The exponents $\alpha$ and $\beta$ are variable model parameters. They
determine the geometry and radial density distribution. We consider
a pillbox disk ($\alpha=0$), a conical disk ($\alpha=1$) or a flared 
disk ($\alpha=9/7$), see also Fig.~\ref{7_geom}. The flaring exponent
of 9/7 is chosen following models for young stellar objects
\citep{chia97}, since there are no better constraints
for disks around active galactic nuclei. $\beta$ determines the
surface density distribution, i.e. the mass concentration with radius. 
We constructed models with a flat mass distribution
($\beta=0$), which implies that the mass is concentrated in the
outer regions for the conical and flaring disk.
In addition, we have models where the mass is concentrated in the
inner regions, either 
similar amounts in logarithmic intervals ($\beta=-2$), or
less centrally concentrated ($\beta=-1$).

The radial optical depth of the disk at the equator is defined at 
20\um\ ($\tau_{20}$), and four values
are used: 5, 10, 20 and 50. A value below 1 at 20\um\ is inconsistent
with observed anisotropies in the mid-infrared \citep{ivb98,ivb00,heck92}. 
The upper value is 
limited by the numerical possibilities of {\sc radical}.
$\Sigma_0$ is chosen such that the model has the desired radial
equatorial optical depth at 20\um.

The inner radius ($r_0$) is either 3 or 30\,pc, the 
outer radius ($r_{\rm out}$) 30 or 300\,pc, which results
in three different sizes for the disk. These will be referred to
as large disk (L\,$=3-300$\,pc), inner disk (I\,$=3-30$\,pc) and outer
disk (O\,$=30-300$\,pc).

\subsection{Dust characteristics}
\begin{figure}[!Ht]
\resizebox{8.7cm}{!}{\includegraphics{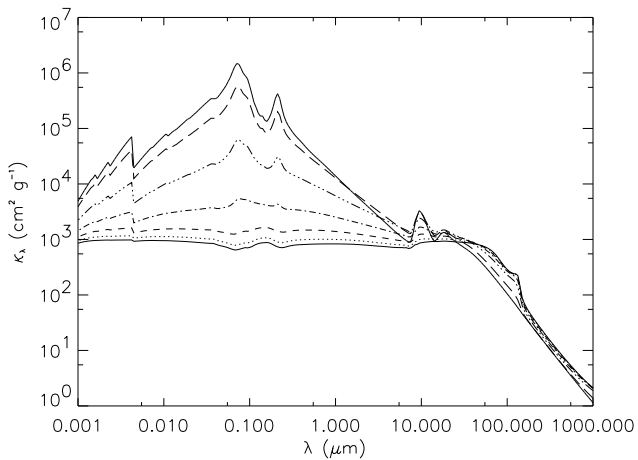}}
\caption{\label{7_opacs}
The resulting opacities for different values of $q$, the exponent
of the grain size distribution. From top to
bottom $q$ increases from --3.5 to --0.5 with steps of 0.5. For the
models we used $q=-2$ (dash-dotted line). The frequency resolution
of this plot is higher than for the models.
}
\end{figure}

Using the extinction efficiency tables from \citet{laor93}
we have constructed an input opacity table. The mass absorption 
coefficient $\kappa_{\lambda}$ is obtained by summing the values 
of $\kappa_{\lambda}(a)$ for all separate grain sizes and dust species, 
using a normalized size distribution $S_i(a)$ for each dust species. 
The normalized size distribution is given by:
\begin{displaymath}
S_i(a) = \frac{a^q\ m_{d,i}\ X_i}{\sum_{a} S_i(a)}
\end{displaymath}
where $a$ is the grain radius, $m_{d,i}$ the mass of a dust grain
of dust species $i$, $X_i$ the fraction of dust species $i$ and
$q$ a dimensionless exponent that can vary between $-0.5$ and $-3.5$. 
In Fig.~\ref{7_opacs} the effect of varying $q$ on the 
resulting opacity is shown. We use only graphite and silicate dust in
a 50/50\% ratio ($X_i=0.5$). The grain sizes range from 0.001 to 10\um.

For $q =-3.5$ the
Standard Galactic Dust or MRN dust is obtained, as described by
\citet{mrn77}. We have made
models with $q=-3.5$ and $q=-2.0$ to test the effect of the
grain size distribution on a conical disk ($\alpha=1$) with a
constant density ($\beta=0$). No silicate 10\um\ emission 
feature has yet been detected in any active galaxy, 
constituting a major problem for radiative transfer models
of obscuring tori so far. The lack of 10\um\ emission might
indicate
that dust in active galaxies is dominated by larger grains than 
Galactic dust \citep{laor93}. Since other studies point in the
same direction \citep{maio01_1}, we have adopted $q=-2.0$ as standard.

\subsection{Input spectrum}
The input spectrum is defined from 10$^{11}$ to 10$^{17}$\,Hz 
i.e. from 3\,mm to 3\,nm (400\,eV), and
divided in 112 equally large logarithmic spacings. This number
was chosen to optimize both calculation speed and frequency 
resolution. The upper limit for the frequency lies at the point
where the dust opacity drops precipitously and the exact value
is of little consequence.
The lower limit lies far enough in the millimeter to 
ensure complete coverage of the infrared output spectrum.

\begin{figure*}[t]
\resizebox{17.cm}{!}{\includegraphics{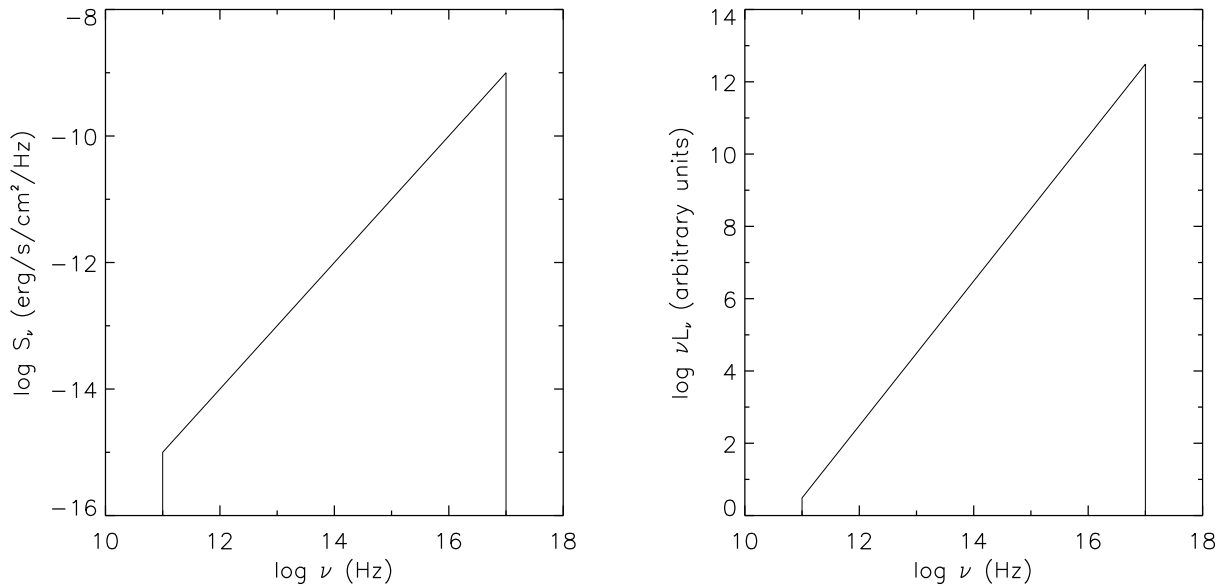}}
\caption{\label{7_input_spec}
The input spectrum used for the models. On the left the flux as it
is used by {\sc radical}, on the right the matching energy spectrum,
using $j=2$.}
\end{figure*}

For the shape of the input spectrum a power-law is assumed:
\begin{displaymath}
L(\nu) = L_0 \nu^j
\end{displaymath}
This allows for two more variables: the spectral slope $j$ and the
total luminosity. The results do not depend strongly on $j$
when it is varied between 0 and 4; we will adopt $j=2$. The 
luminosity is scaled to be 10$^{12}$\,L$_{\odot}$, integrated
over the highest energies (100--400\,eV), 
which is a typical intrinsic X-ray luminosity for radio-loud 
active galaxies. In Fig.~\ref{7_input_spec} we show the spectrum
that we use as input for {\sc radical}.

\subsection{Grid setup}
The radial grid is set up using 
$r_0$ and $r_{\rm out}$. A fixed number of radial points is 
defined
between the inner and outer radius. The grid is refined at the
inner edge to prevent flux overshoots in the innermost grid cells.
For the small disks, we have used 40 radial gridpoints; for the large
disk 60. 
The angular grid does not need to extend from pole to pole. The
dust density in the wings of the Gaussian is negligible at
a few scale heights, and the disk can be mirrored in the
equator. The grid is therefore defined from the equator to a maximum
angle that extends well beyond the height where the opacity
reaches unity. The code
assumes that the remaining space has the same density as the
grid cells on the edge of the defined angular raster. We use
20 gridpoints in angular direction for all models. The resulting
grid is depicted in Fig.~\ref{7_geom}.

\begin{figure*}[!Ht]
\resizebox{8.7cm}{!}{\includegraphics{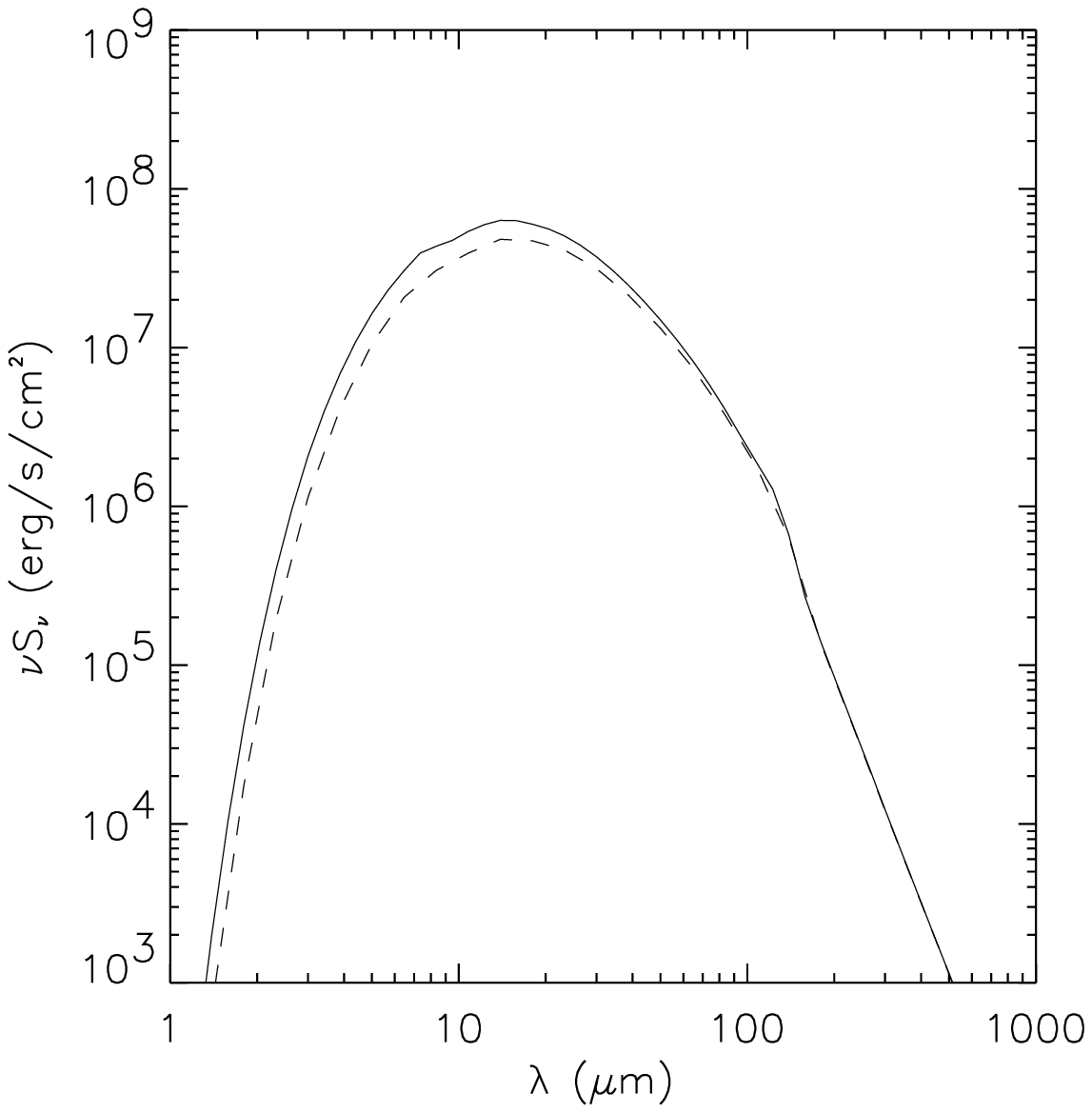}}
\resizebox{8.7cm}{!}{\includegraphics{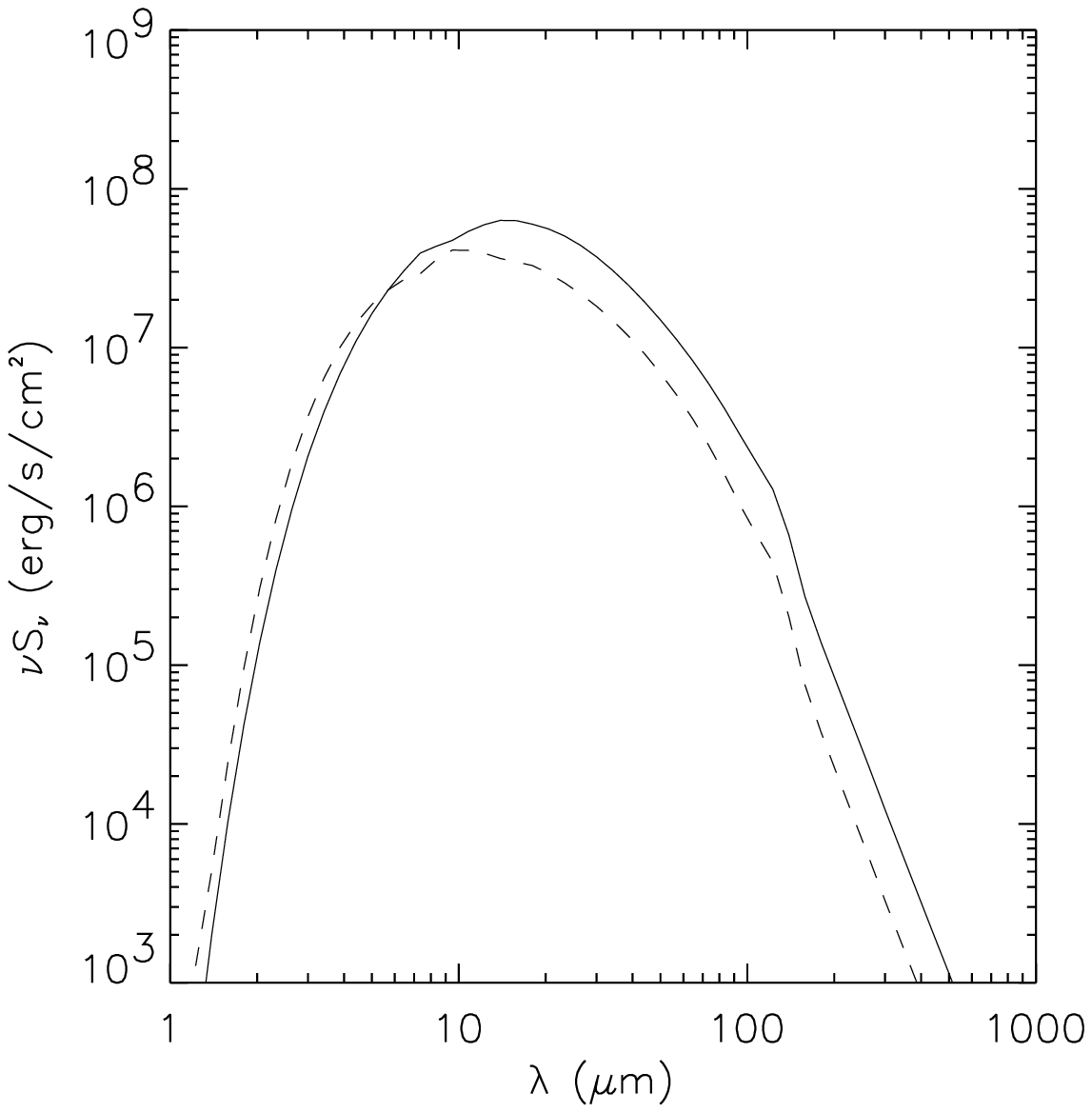}}
\caption{\label{7_low_h0}
{\it Left panel}: reference model compared with
a model that has lower frequency resolution (dashed line).
{\it Right panel}: comparison of the reference model ($h_0=3$\,pc,
solid line) with a model that has a lower value for the 
inner scale height ($h_0=1$\,pc, dashed line). 
}
\end{figure*}

\section{The radiative transfer code {\sc radical}}
\label{7_radical}
In order to compute the spectrum emitted by the irradiated torus, 
a 2-D radiative transfer calculation is performed. This is done using 
{\sc radical}\footnote{For more information, see also
http://www.mpa-garching.mpg.de/PUBLICATIONS/DATA/radtrans/radical/}, 
which is an all-purpose 2-D radiative transfer code for
axisymmetric circumstellar media \citep{dul00}. In this
section we describe how we use this code to solve the transfer problem 
of a torus, partially obscuring the active nucleus.

The dust in the torus is assumed to be in equilibrium with the radiation
field. Each dust grain has a fine-tuned temperature such that it emits as
much energy per second as it absorbs. In the case of an optically thin torus
this temperature is easily computed, since the flux of the nucleus is known
and simply goes as $1/r^2$. In such a case, no radiative transfer is needed,
since the temperatures of the grains can be computed. A simple
volume integral over the emission is then sufficient to obtain the 
emitted spectrum.

Observations of radio-loud active galaxies suggest that the
opacity of the obscuring disk exceeds unity at least up to 
mid-infrared wavelengths ($\sim$40\um). The calculation of the emission 
changes significantly in optically thick models.
The nuclear radiation is now stopped when it is intercepted by the
torus at the inner edge, and slowly diffuses into the dust. The inner 
edge is therefore expected to be very 
hot, while the interior of the torus is much
cooler. The spectral energy distribution of such a torus with a large range
in temperatures will be broader than expected from an optically
thin torus. Solving the temperature of the dust grains is much more
difficult than in the optically thin case. This temperature depends not
only on the incident nuclear flux, but also on the infrared radiation of the
torus itself. A grain located many optical depths deep into the torus will
see little of the direct nuclear radiation, but instead will see the
infrared emission of its neighboring grains. The radiative equilibrium of
this grain depends therefore on the emission of the grains elsewhere in the
torus. Every element of the torus is radiatively connected to every other
part.

\subsection{The transfer algorithm}
At present, the best way to solve this problem is to perform
a 2-D radiative transfer calculation. The absorption and 
re-emission of radiative energy will be modeled as it seeps 
through the torus. The radiative transfer along each ray 
through the torus is given by:
\begin{equation}\label{7_eq-trans}
\frac{dI_\nu}{ds} = \rho \kappa_\nu (B_\nu(T)-I_\nu)
\end{equation}
where $B_\nu(T)$ is the Planck function at the local temperature, 
$I_\nu$ the intensity, and $s$ the path length along the ray. 
The solution of the radiative transfer proceeds as follows. We start with an
initial guess for the dust temperature distribution. Using a radiative
transfer integrator based on the algorithm of {\em Extended Short
Characteristics} \citep[see][for details]{dul00}, the radiation
field at every location in the disk is computed as a function of frequency
$\nu$ and direction $\mu,\phi$ of the radiation. 

From this radiation field the mean intensity $J_\nu$ is computed:
\begin{equation}
J_\nu \equiv\frac{1}{2}\int I_\nu d\Omega
\end{equation}
where $d\Omega$ is the solid angle. The dust temperature at every point can
then be found by solving the equation:
\begin{equation}
\int_0^\infty B_\nu(T)d\nu=\int_0^\infty J_\nu d\nu
\end{equation}
Once this (new) temperature structure is found, it is compared to the
old one. If the relative difference is larger than 0.001, we go back to
the integration of the transfer equation (Eq.\,\ref{7_eq-trans}), and keep
iterating until the relative successive difference drops below the 
convergence criterion of 0.001. We tested the convergence by starting
the model with a cold (1\,K) and a hot (1500\,K) disk and comparing
the resulting spectra. The results are indistinguishable, which
indicates that the models are fully converged.

\begin{figure*}[!Ht]
\resizebox{8.7cm}{!}{\includegraphics{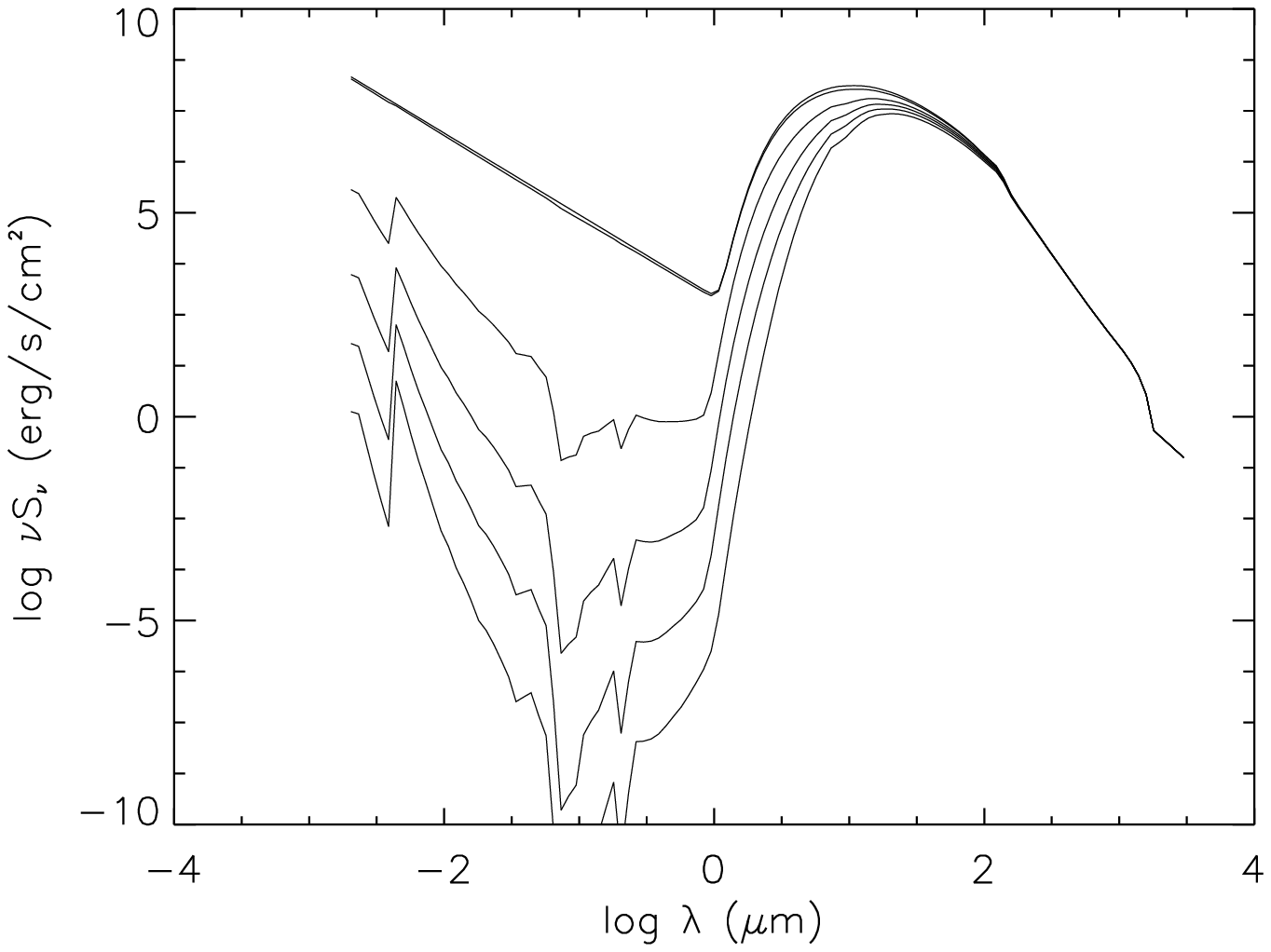}}
\resizebox{8.7cm}{!}{\includegraphics{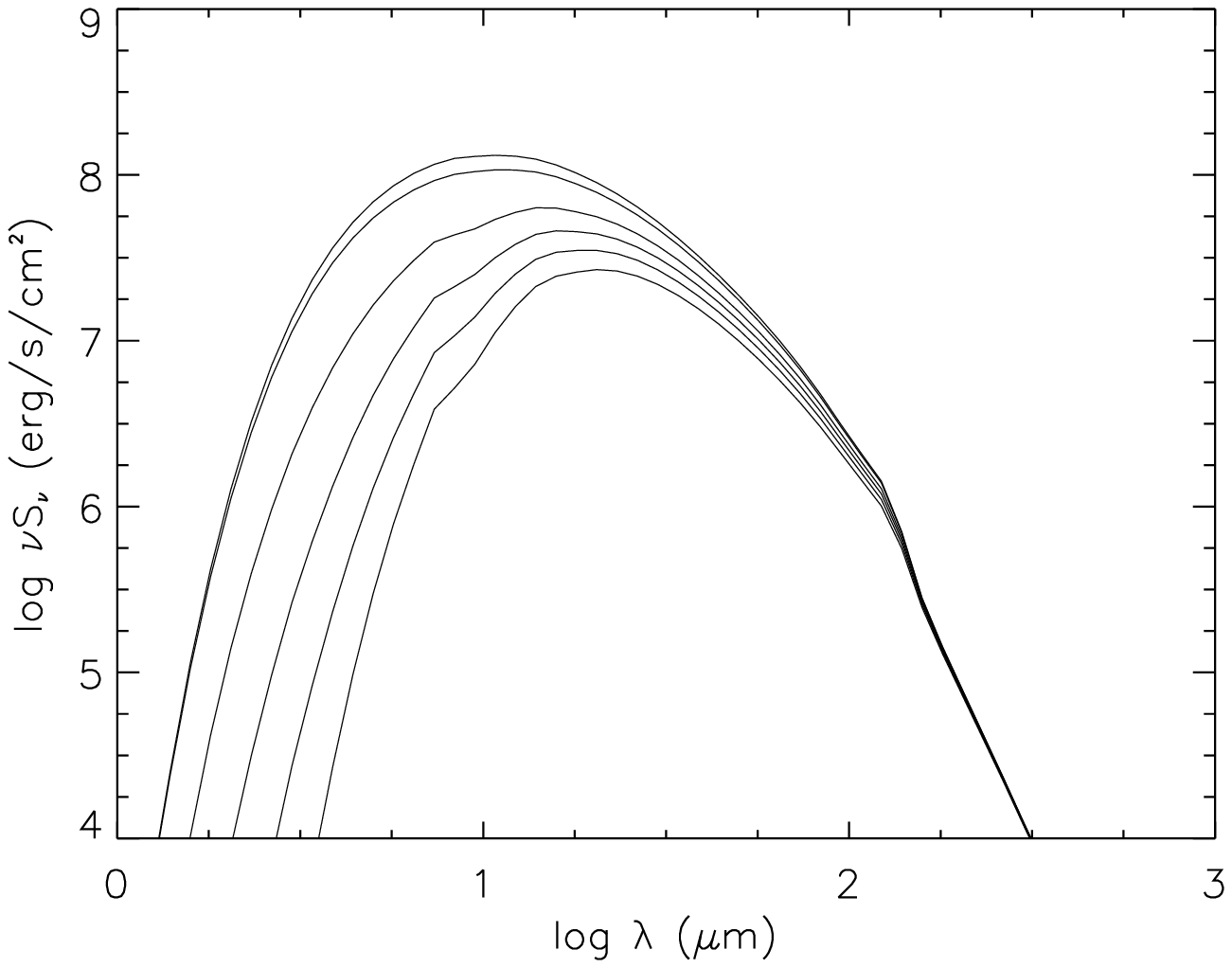}}
\caption{\label{7_ref_model}
The reference model, using $q=-2$, $\alpha=1$, $\beta=0$, $\tau_{20}=10$, 
with a disk size from 3 to 30\,pc. Different lines are for different 
aspect angles, the top line corresponding to $\theta=5^{\circ}$, next 
are 30, 50, 60, 70 and $90^{\circ}$. In the right panel a zoom on the 
infrared regime is displayed.}
\end{figure*}

\subsection{Computing the spectrum}
The spectrum is obtained from the density and temperature distribution 
using a ray-tracing algorithm. For each discrete frequency an image of 
the torus is produced. These images are then integrated 
to obtain the spectrum
$F_\nu$. To make sure that the pixels of these images accurately sample all
relevant scales in the system, we distribute the pixels of the image not in
a rectangle (as in normal digital images), but in concentric circles around
the central source. Each circle contains 64 pixels, and the circles
themselves have radii such than for every radial gridpoint on the
computational domain there are two circles in the image. In this way there
will be an equal number of pixel-circles per logarithmic interval in
radius, and this guarantees that both the very inner parts of the disk
(few pc) as well as the far outer parts (tens to hundreds of pc) will be 
sampled properly. 

The spectra can be computed at any requested inclination. We
select inclinations ($\theta$) of 5, 30, 50, 60, 70 and 90 
degrees. For small $\theta$ the inclination cannot be smaller
than 0.1 radian, due to limitations of the code, so its lowest 
value is 5.7 degrees. All 2-D/3-D
effects are included. For edge-on inclination ($\theta=90^{\circ}$) 
one will see only the outer edge of the disk, while at more face-on 
inclination (say $\theta=5^{\circ}$) one will also see the central 
source and the warm inner regions of the disk. The angle at which the
inner regions are shielded from view depends on the opening angle
of the torus and the optical depth in our models.

\subsection{Flux conservation and resolution tests}
In the ideal situation, the torus will emit exactly the same amount 
of luminosity
as it has absorbed from the central source. This requirement of flux
conservation is a powerful check on the results of the code, and this check is
in fact always performed by {\sc radical} at the end of each run. Moreover,
{\sc radical} is built in such a way that radiation in optically thin areas
of the system does not loose or gain flux through numerical diffusion.
Unfortunately, in areas of large optical depth, flux conservation is not
intrinsically guaranteed by the code, and occasionally the flux can be
slightly over-predicted. In most cases we have energy conservation to within
about 10\%, but sometimes (larger disk and/or high opacities) this error 
can go up to 20\%. Refinement of the
grid and fine-tuning of the boundaries of the computational domain usually
help to reduce this error, but there are feasibility limits to the number of
grid-points that can be used. In this paper all models stay within
this 20\% tolerance, and most of the models in fact stay within $\sim$10\%. 

The frequency resolution has to be high enough to prevent flux
losses or overshoots. In Fig.~\ref{7_low_h0} the effect of a lower 
frequency resolution is shown: with half the number of frequency 
points the flux deviates less than a factor of two. A higher frequency 
resolution does not significantly change the results.

Changing the scale height at the inner radius also affects the
output spectrum. As shown in Fig.~\ref{7_low_h0},
decreasing $h_0$ mainly results in a lower output flux in the 
far-infrared. Since the disk is thinner, there
is less cool dust in the disk. The decrease
of flux is proportional to the decrease in $h_0$.
It only marginally affects the shape of the spectrum, therefore it 
can be used to tune the results to observations for individual objects. 
When comparing colours, the assumed value of $h_0$ is irrelevant.

\begin{figure*}[!t]
\begin{center}
\resizebox{8.2cm}{!}{\includegraphics{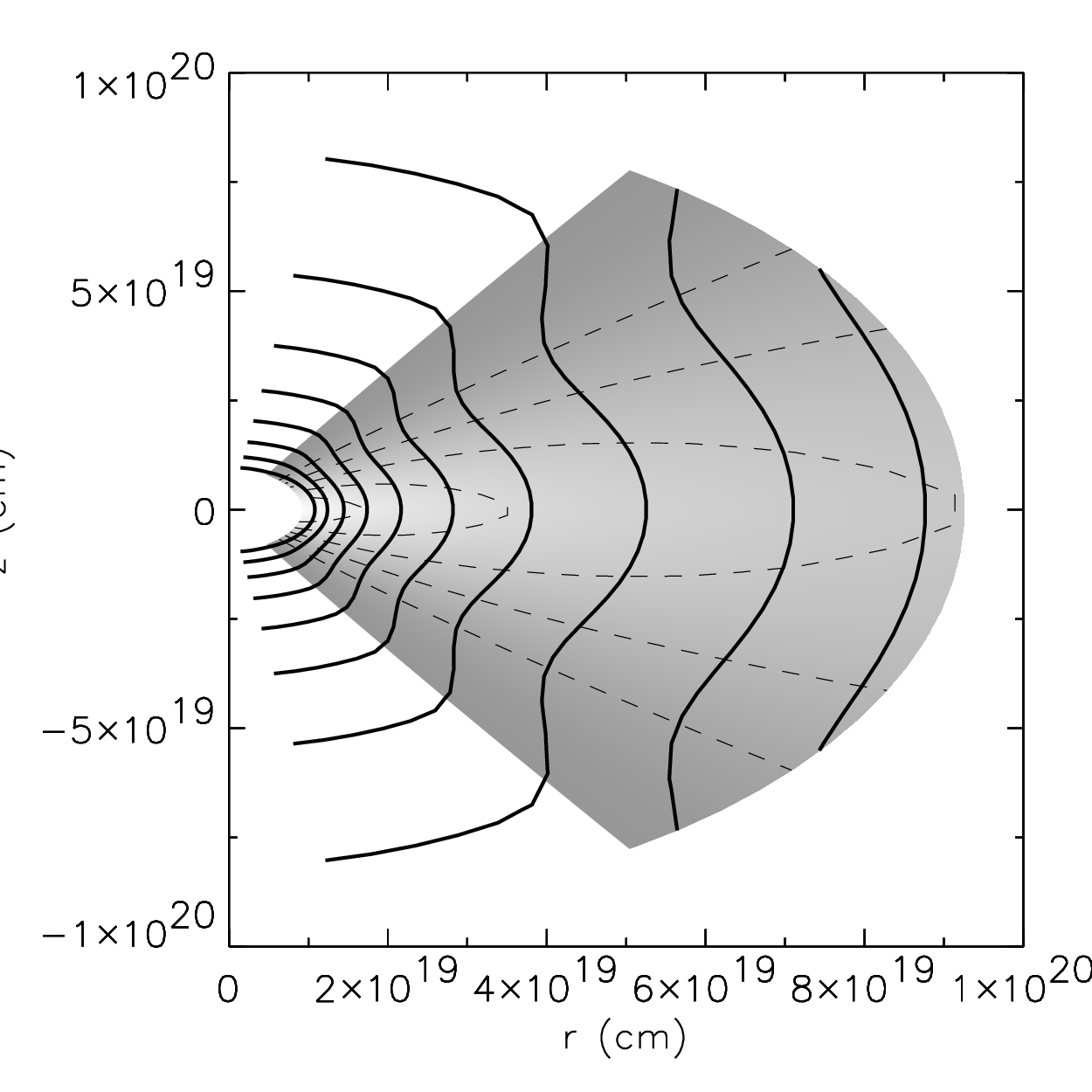}}
\resizebox{8.2cm}{!}{\includegraphics{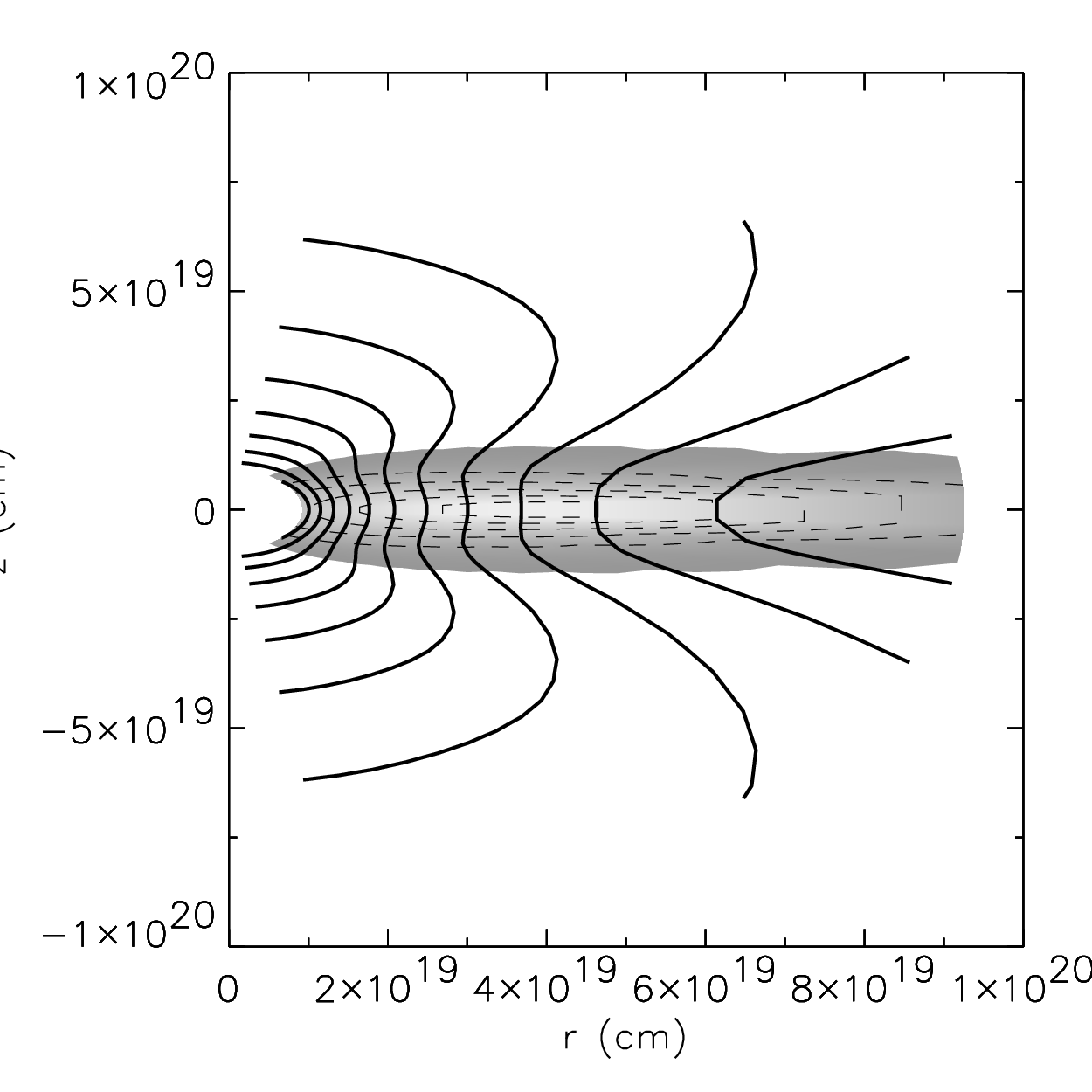}}
\resizebox{8.2cm}{!}{\includegraphics{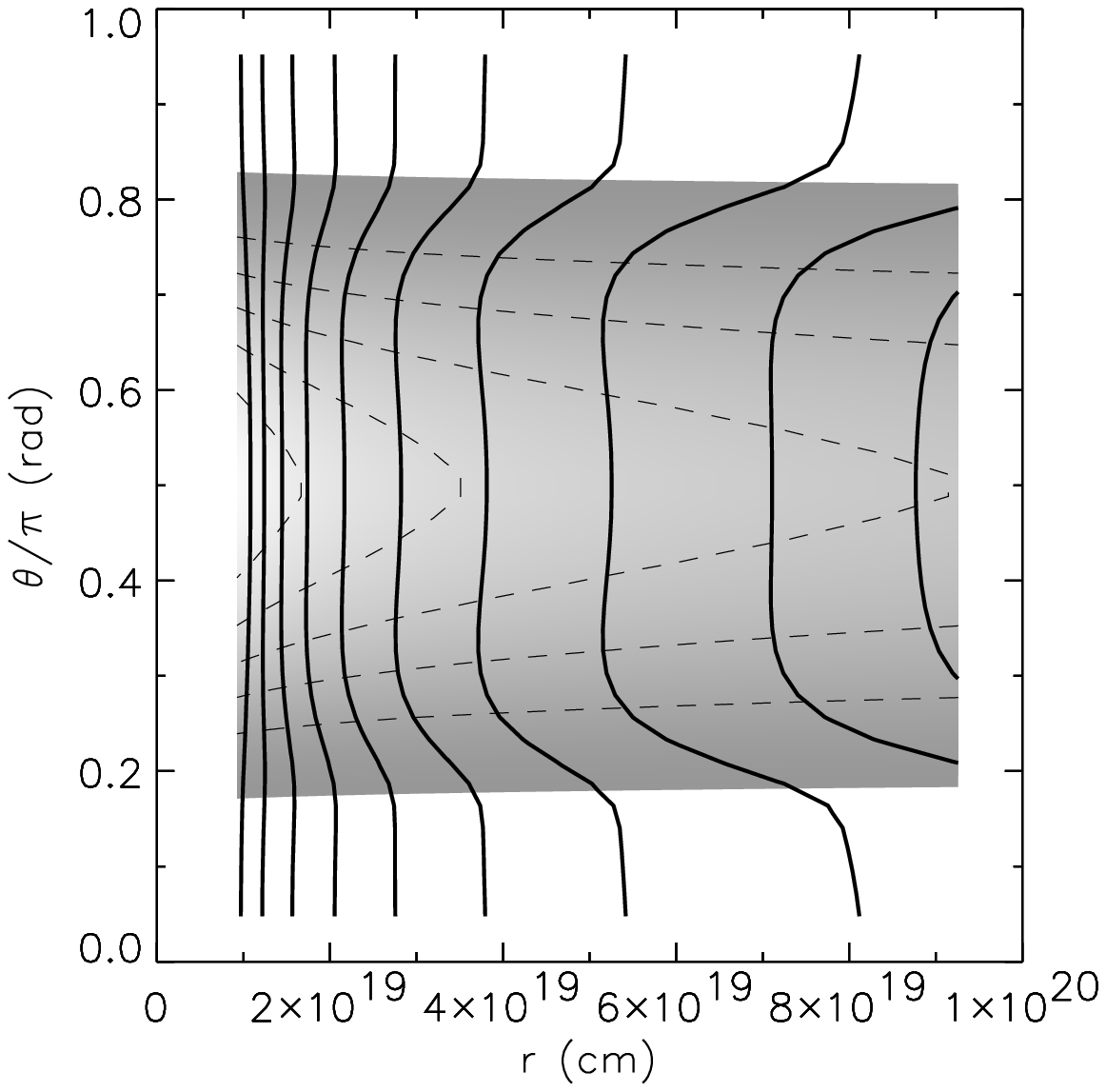}}
\resizebox{8.2cm}{!}{\includegraphics{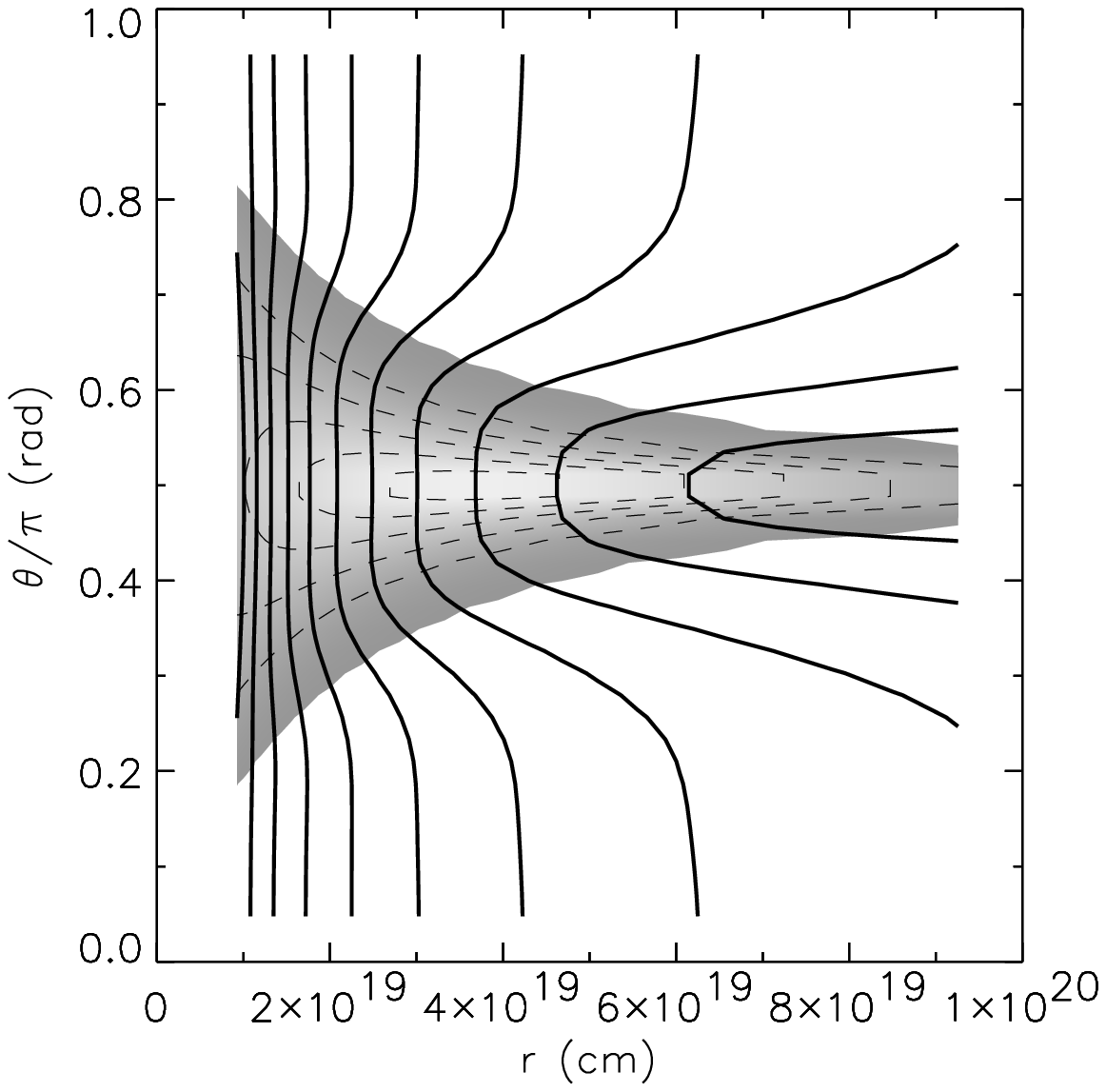}}
\caption{\label{7_temp_dens_ref}
The temperature and density distribution for the conical disk 
(left two panels) and the pillbox disk (right two panels).
Upper panels are in $z$ coordinates, lower panels in $\theta / \pi$.
Grey scale shows the density distribution of the dust, the
solid lines are temperature contours, corresponding to
600\,K (leftmost) and decreasing with steps of 50\,K.
}
\end{center}
\end{figure*}

\subsection{Improvement on previous methods}
The large parameter space probed by our models is unique and
allows for a full comparison of observations with these models.
The dependence of the output spectrum on disk 
parameters has not been studied in this much detail before.
For simplicity we have chosen to deal with a smooth dust
distribution, before assessing the effects of a clumpy
medium.
One of the main problems has always been to reproduce the
broad infrared SED observed in active galaxies: the effects
of the geometry of the torus, its density, size, or opacity are not
clear.  We compare the models to available broad band data in
order to assess the width of the resulting spectrum. Instead of
fitting a large sample of objects, we will compare the colours of
our models to the observed colours. Only for \cyga, a powerful
nearby radio galaxy, will we study the actual fit in detail. 

A second important issue has been the occurence of 10\um\
emission at small inclinations in most literature models, which is 
not observed. Although our
frequency resolution does not allow an in depth study of the
behaviour of the 10\um\ feature, we will discuss the effects
of altering the grain size distribution on our models. The
powerful radio-loud active galaxies lack proper observations
of the 10\um\ region, and since we mainly aim to understand 
these objects, no direct comparison to infrared data will be 
presented in this paper. In a consecutive paper, high resolution
models will be discussed, allowing for a direct comparison with
new observations from the SIRTF satellite for a range of active
galaxy types.

\begin{figure*}[!Ht]
\resizebox{17.cm}{!}{\includegraphics{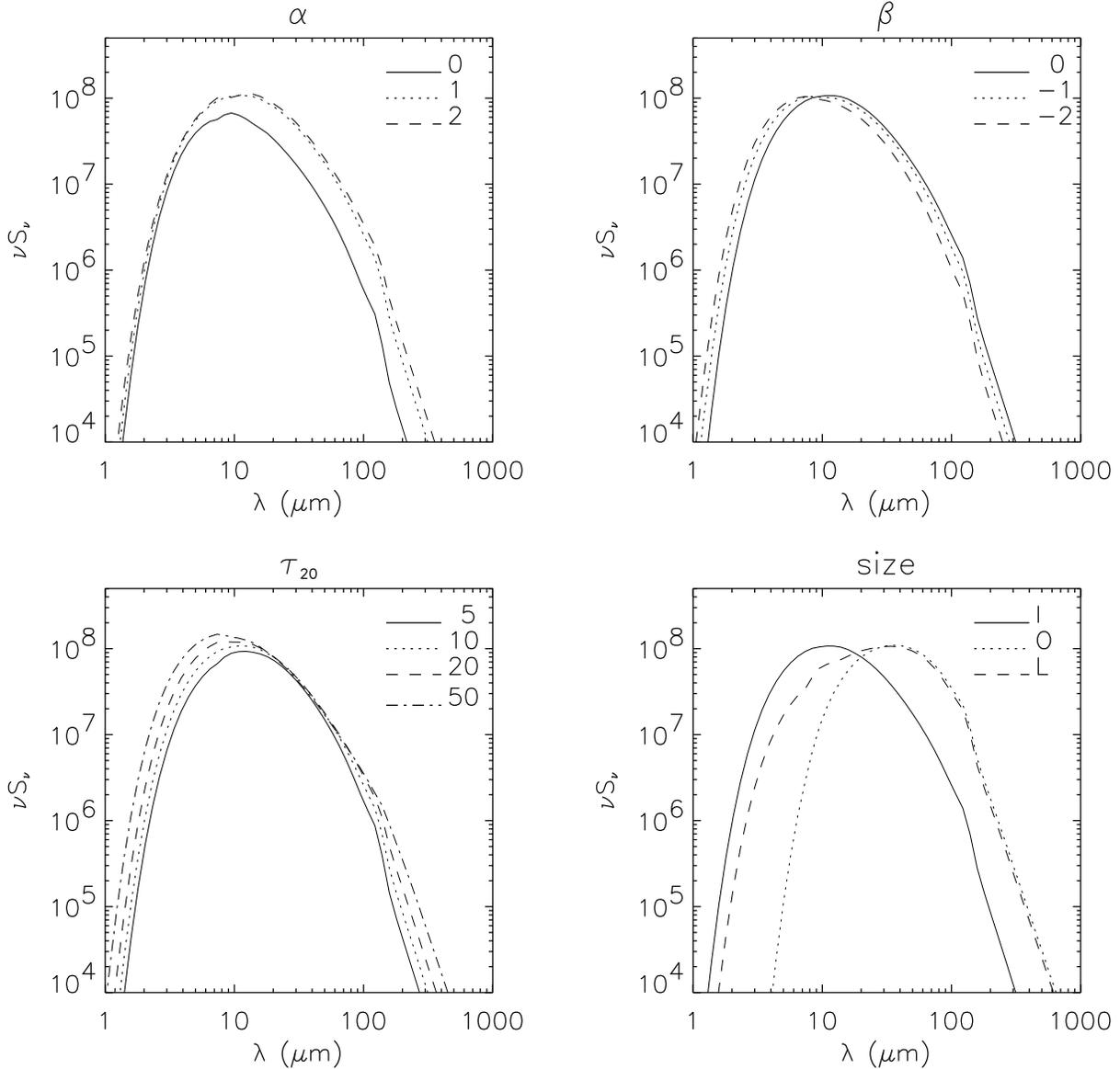}}
\caption{\label{7_model_compare30}
Comparison of the models with different varying parameters for
a fixed inclination of $\theta=30^{\circ}$. The non-varying parameters 
have the values as for the reference model: $\alpha=1$, $\beta=0$,
$\tau_{20}=10$ and size 3--30\,pc.}
\end{figure*}

\section{Results}
In order to understand the effect of the different variables, 
we have defined a reference model, to which all other models 
are compared. The relevant parameters are given in 
Table~\ref{7_model_params}.
The reference model is chosen to be a conical
disk with constant surface density, a size from 3 to 30\,pc and
optical depth of 10. This will allow a direct comparison
to literature models, which generally have similar characteristics.
However, we will use $q=-2.0$, which does introduce a different
behaviour when considering the 10\um\ silicate feature, but which
is a more realistic value when considering active galaxies.
The optical depth of 10 is chosen to account for observed 
differences at 40\um\ restframe wavelength in quasars and 
radio galaxies \citep{heck92,ivb98,ivb00}. The size of
the disk corresponds to sizes found in Seyfert galaxies, although
it may be significantly larger in powerful radio galaxies.

\subsection{The reference model}
The model spectrum is shown in Fig.~\ref{7_ref_model}. 
The spectrum peaks between 10 and 20\um, depending on inclination.
For the different inclinations there are clear flux differences with 
inclinations (anisotropies) up to 100\um. 
The knee at 110\um\ is a dust opacity feature. There
is a slight indication of 10\um\ silicate absorption at larger
angles. 

\begin{figure*}[t]
\resizebox{17.cm}{!}{\includegraphics{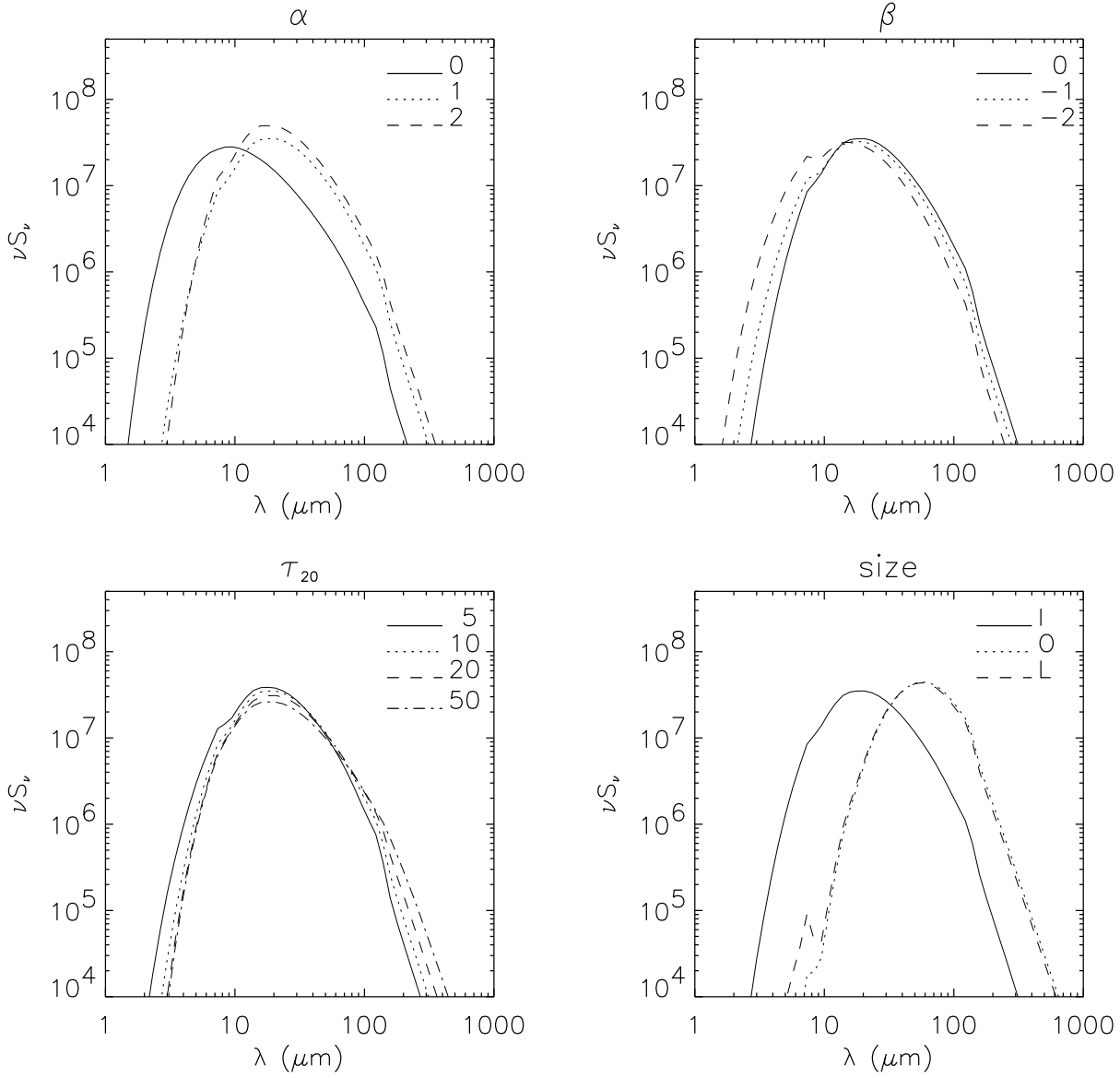}}
\caption{\label{7_model_compare70}
As Fig.~\ref{7_model_compare30}, but for inclination angle
$\theta=70^{\circ}$.}
\end{figure*}

The temperature and density distribution for the reference model
are depicted in Fig.~\ref{7_temp_dens_ref}.
The temperature in the disk is lower than in its 
optically thin surroundings, which is to be expected. In
the inner regions the temperature drops steeply with radius, 
flattening at larger radii. Within the disk, the temperature in angular
direction is more or less constant, with a steep increase at
the edges to reach the temperature of the surrounding medium.

\begin{figure*}[t]
\begin{center}
\resizebox{7.cm}{!}{\includegraphics{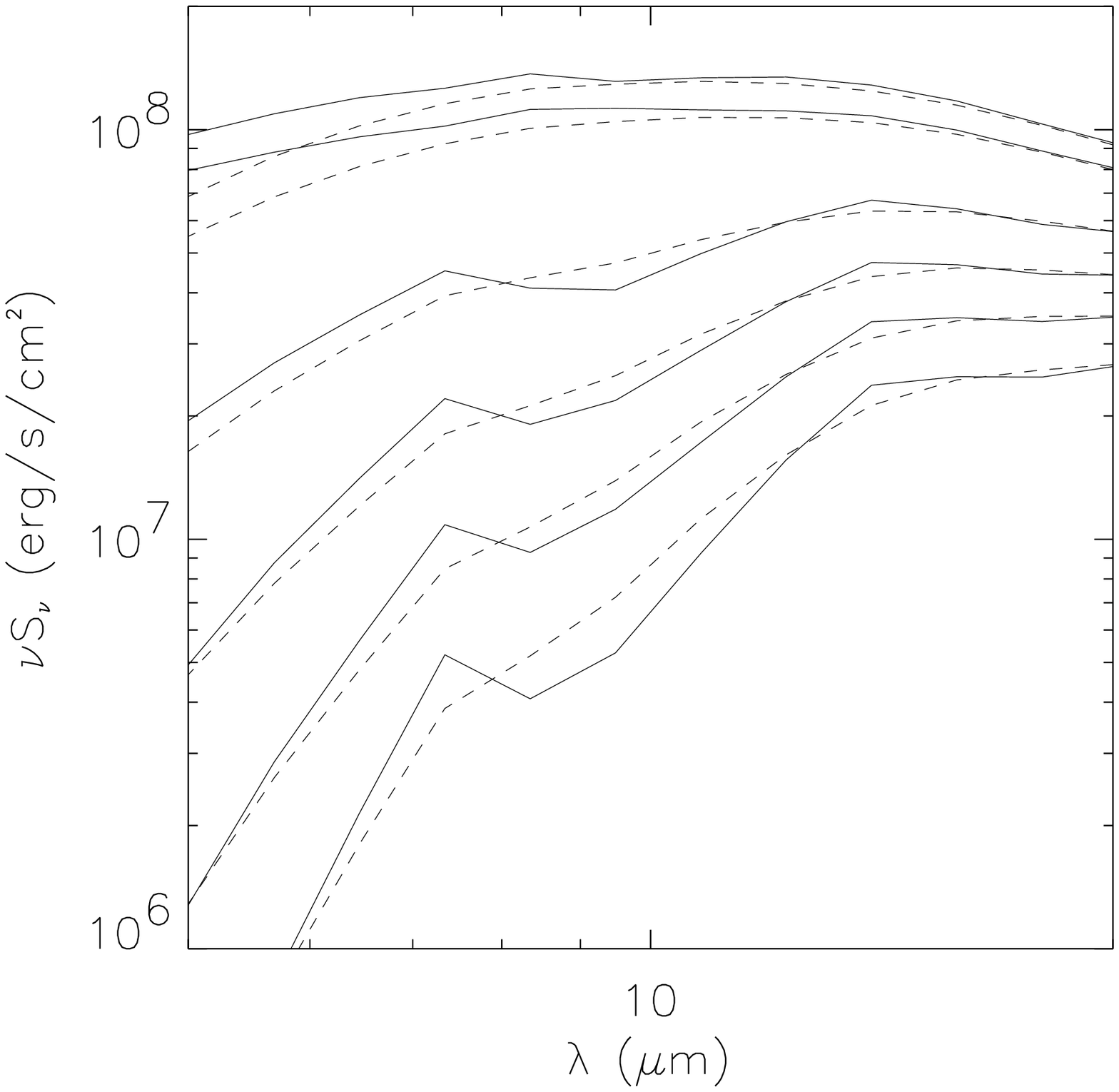}}
\hspace{1cm}
\resizebox{8.cm}{!}{\includegraphics{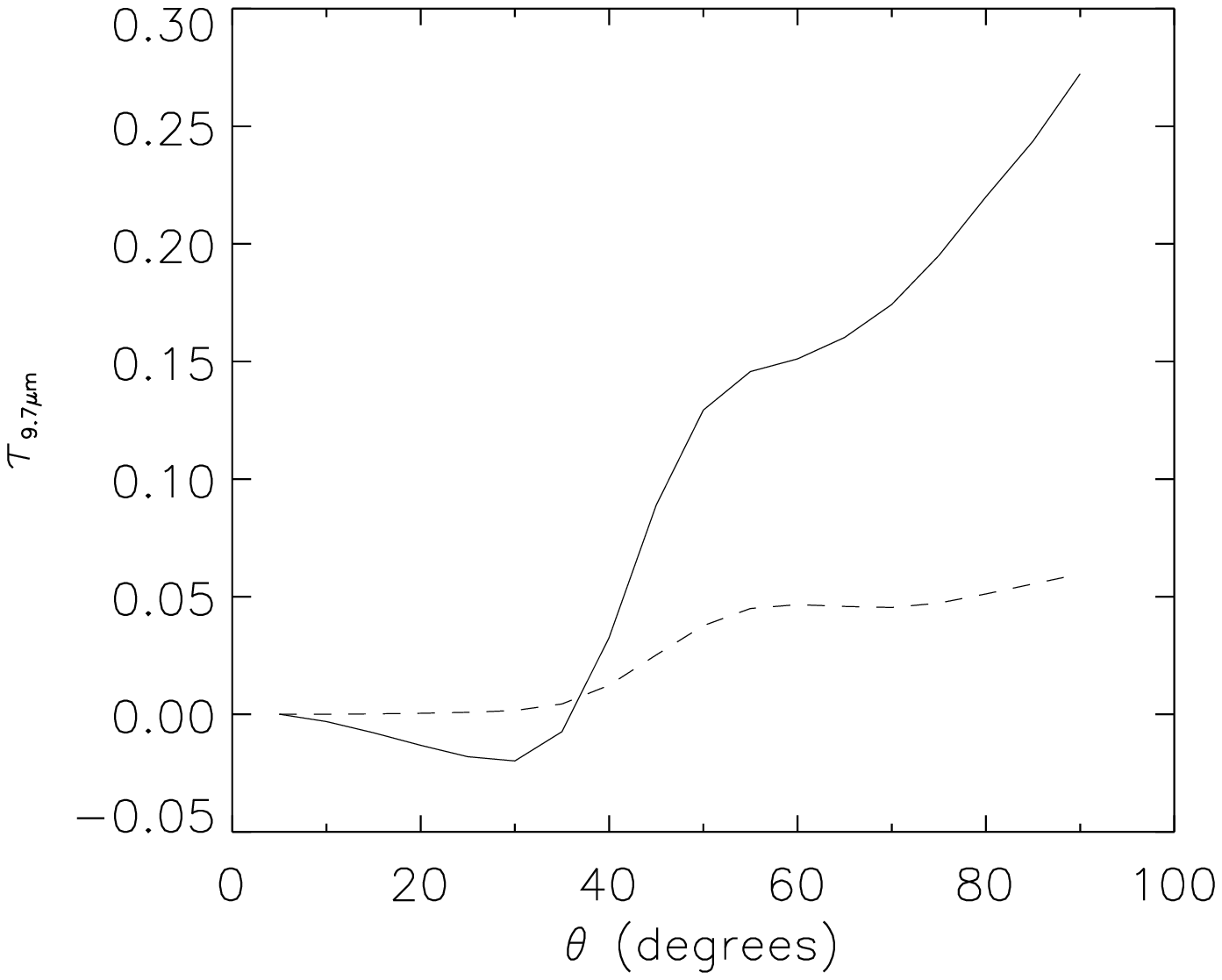}}
\caption{\label{7_q_compare}
Comparison of the 10\um\ silicate absorption in the reference model when 
using different opacity laws. The 
solid line represents MRN dust ($q=-3.5$, Galactic), the dashed line the 
larger grain dust ($q=-2$) adopted in this paper. {\it Left panel}: A zoom
on the spectrum for inclinations 5$^{\circ}$, 30$^{\circ}$, 
50$^{\circ}$, 60$^{\circ}$, 70$^{\circ}$ and 90$^{\circ}$ (from top to
bottom). {\it Right panel}: Depth of the 10\um\ absorption 
feature with inclination (positive values indicate absorption).}
\end{center}
\end{figure*}

Below is described what the effect of varying the different 
parameters has on the model results, which is shown in
Figs.~\ref{7_model_compare30}, \ref{7_model_compare70}
and \ref{7_q_compare}. These figures are a concise representation
of the effects of changing the parameters. A complete
set of hundreds of figures has been used to study the
effect of the variables on anisotropy and 10\um\ absorption 
in the models. 

The depth of the 10\um\ absorption shown in Fig.~\ref{7_q_compare}
is determined using a similar method as \cite{pk92}: a power-law 
is fitted through the 6.5 and 13.9\um\ points. However, since the 
model curves have a grey-body shape, fitting a power-law will mimick silicate 
emission where none is present. This is circumvented by adding to 
the power-law fits the residual from the model with 5$^{\circ}$
inclination, where no silicate emission or absorption occurs, resulting 
in a continuum that approaches a grey-body 
shape in the region 5--15\um. Subsequently, the difference between the 
continuum and the 9.7\um\ gridpoint is measured. The depth $\tau_{9.7\mu m}$ 
is positive for aborption.

\begin{itemize}
\item $\alpha$: changing the physical shape of the disk to
	a pillbox ($\alpha=0$) or flaring disk ($\alpha=9/7$). 
	The flaring disk and conical disk have similar spectra,
	slightly varying with inclination. The pillbox 
	disk is less luminous at all inclinations, but this is 
	more obvious at smaller angles.
	It is hotter than the reference model, with
	the peak at shorter wavelengths for smaller inclinations. 
	The wavelength of
	the emission peak does not shift significantly with
	inclination, but there is an indication of increased anisotropy. 
	The flared models have a higher degree of isotropy than 
	the reference model, but show a similar shift in the 
	peak wavelength with inclination.
	They also show a hint of 10\um\ absorption at small
	inclinations, but this effect depends strongly on
	the optical depth and the size of the disk.

\item $\beta$: changing the surface density from a constant
	to a more centrally concentrated configuration ($\beta=-2$).
	For fixed
	$\alpha,$ the models differ slightly at large inclinations,
	and marginally at small inclinations. The more centrally
	concentrated, the more the spectrum moves to shorter wavelengths. 
	At large $\theta$, there is a hint of 10\um\ absorption, which
	deepens when $\beta$ changes from 0 to --2. The depth does
	not depend on the value of $\tau_{20}$. The
	10\um\ absorption is absent in the 30--300\,pc disk.

\item $\tau_{20}$: changing the optical depth at 20\um\ from 5 to 50.
	The higher opacity models have broader infrared spectra
	and shorter peak wavelengths at small angles. For larger 
	inclination the high opacity models are not much broader, 
	but clearly have more far-infrared emission. The
	peak wavelength hardly changes at larger angles.
	An important effect is the increase of anisotropy
	with increasing $\tau_{20}$, but the wavelength at which	
	the dust becomes optically thin does not vary significantly.
	There is a hint of 10\um\ absorption
	in the low opacity models at large inclination.

\item size: the outer disk emits at much longer wavelengths than 
	the inner
	disk, and shows a stronger anisotropy at longer
	wavelengths. The large disk has a much broader
	spectrum at small angles, indicating a larger temperature 
	range. At large inclinations the outer and the large
	disk are almost identical. The large disk has clear 10\um\ 
	absorption. A hint of this is seen in the inner and outer disk.

\item $q$: using Galactic dust instead of a large-grain dominated
	mixture. For larger inclinations the 10\um\ absorption
	becomes deeper. There is a hint of 10\um\ emission,
	which strongly depends on the optical depth and disk
	size. For smaller optical depth, the emission is clearly
	present. For larger optical depth, the absorption is
	only seen in intermediate inclinations. A direct 
	comparison of $\tau_{9.7\mu m}$ shows minor silicate 
	emission at inclinations around 30$^{\circ}$, but up to
	a factor of 5 increase of silicate absorption at angles 
	larger than 40$^{\circ}$. In the outer
	disk, the absorption is deeper at larger inclinations,
	but there is a clear emission feature for small angles.
	The anisotropy is less pronounced than in the reference
	model, although it is visible up to 100\um\ as well.
	The peak wavelength is unchanged.
\end{itemize}

\begin{figure*}[t]
\begin{center}
\resizebox{6.8cm}{!}{\includegraphics{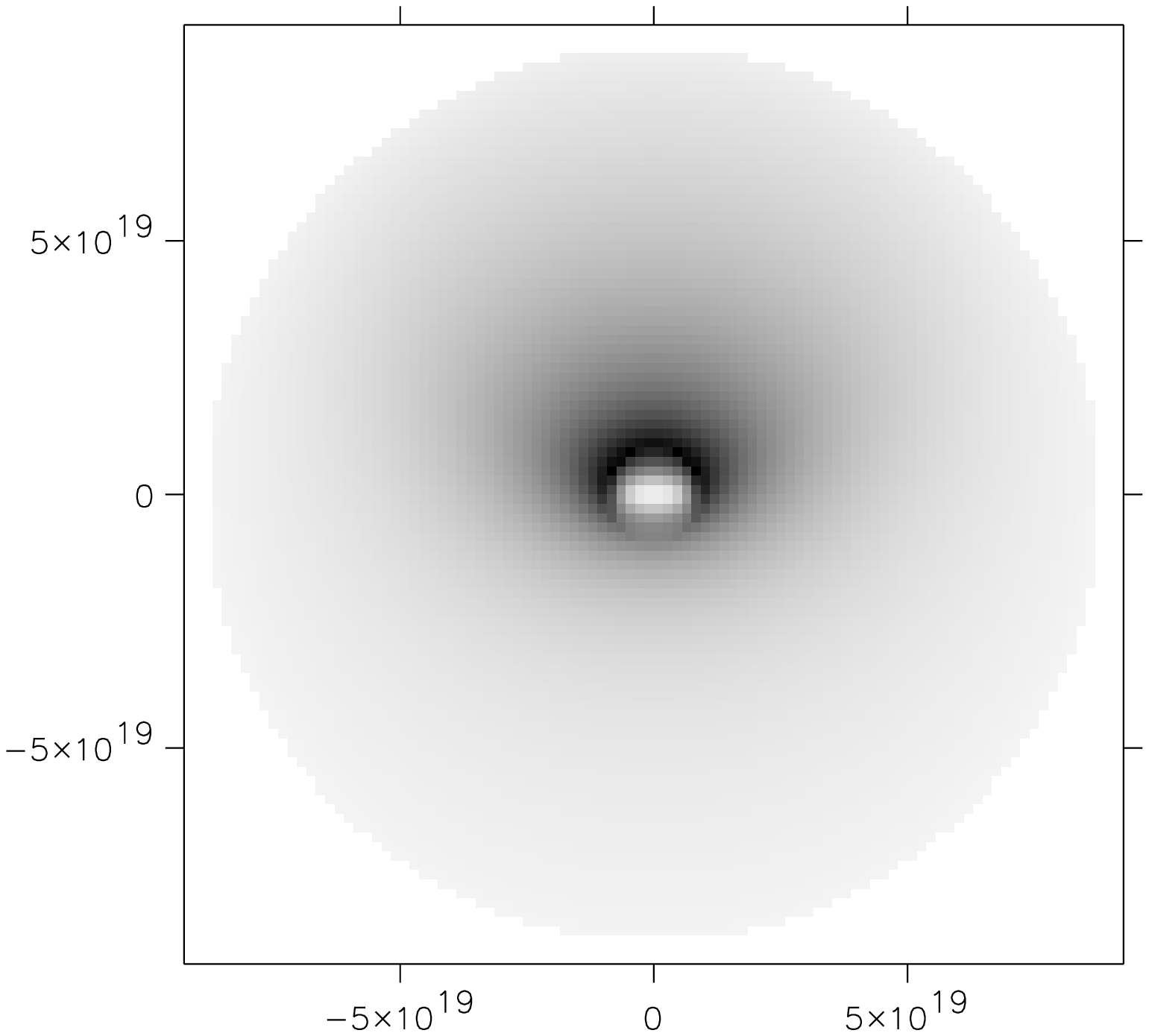}}
\resizebox{6.8cm}{!}{\includegraphics{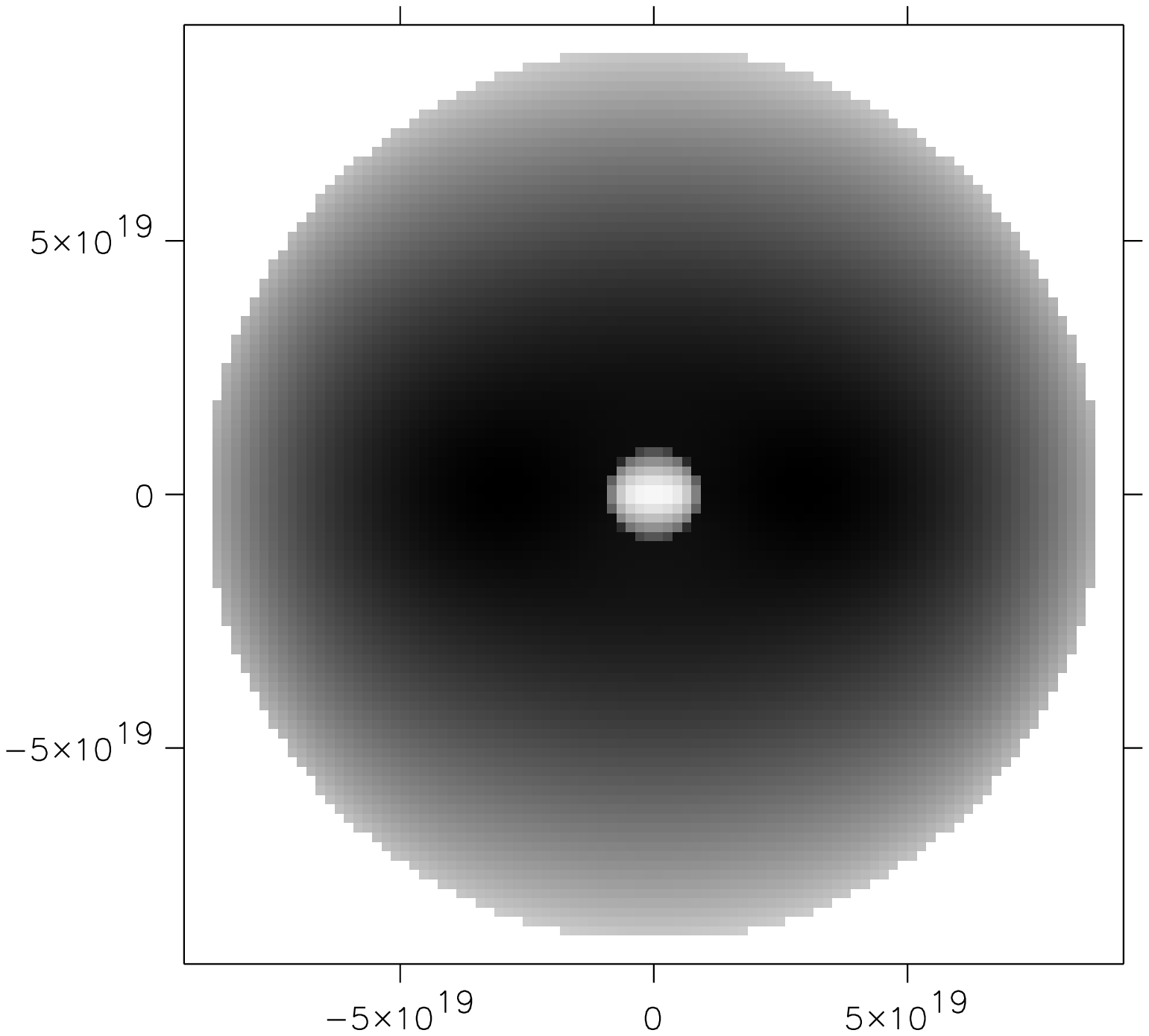}}
\hfill

\resizebox{6.8cm}{!}{\includegraphics{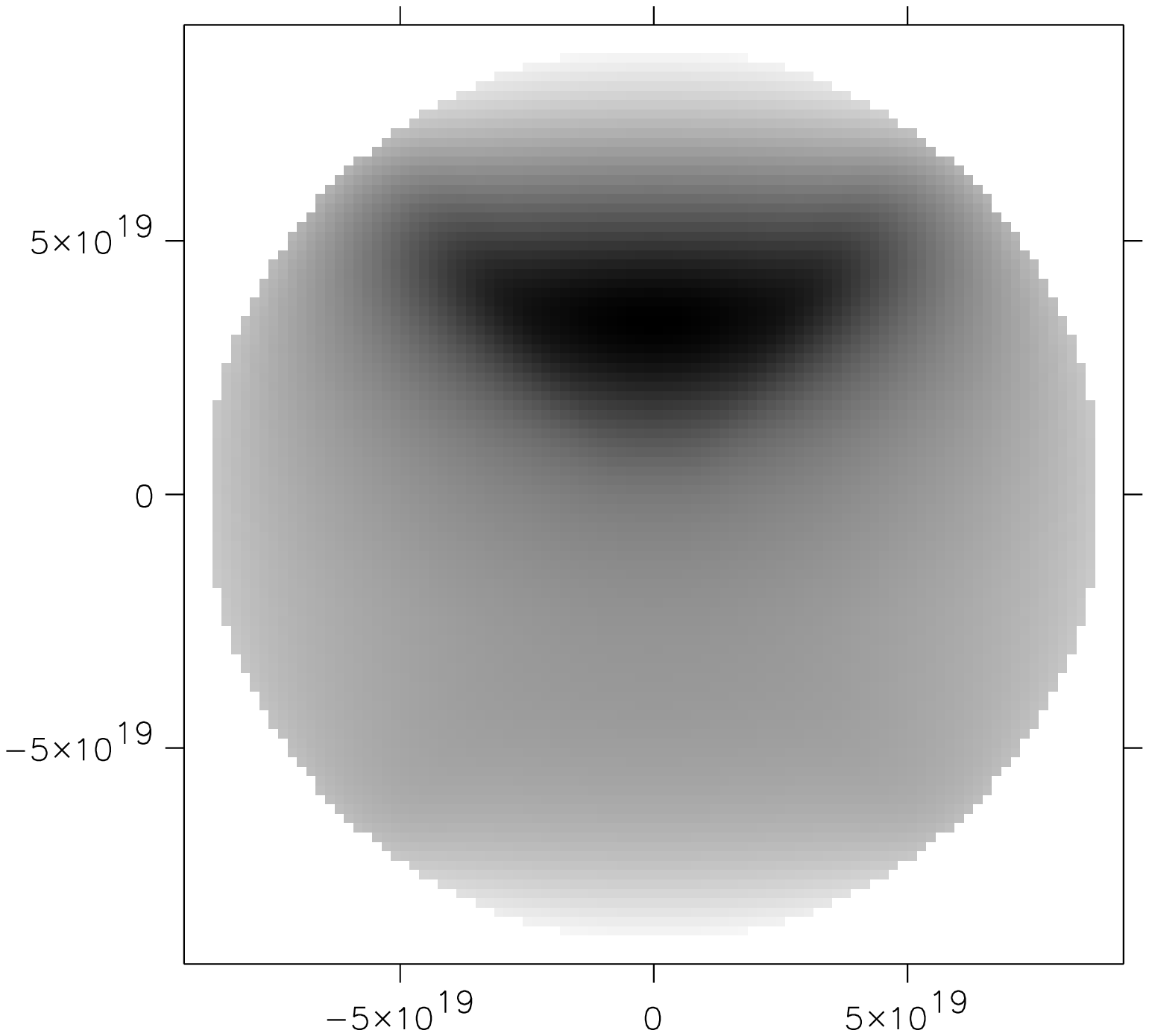}}
\resizebox{6.8cm}{!}{\includegraphics{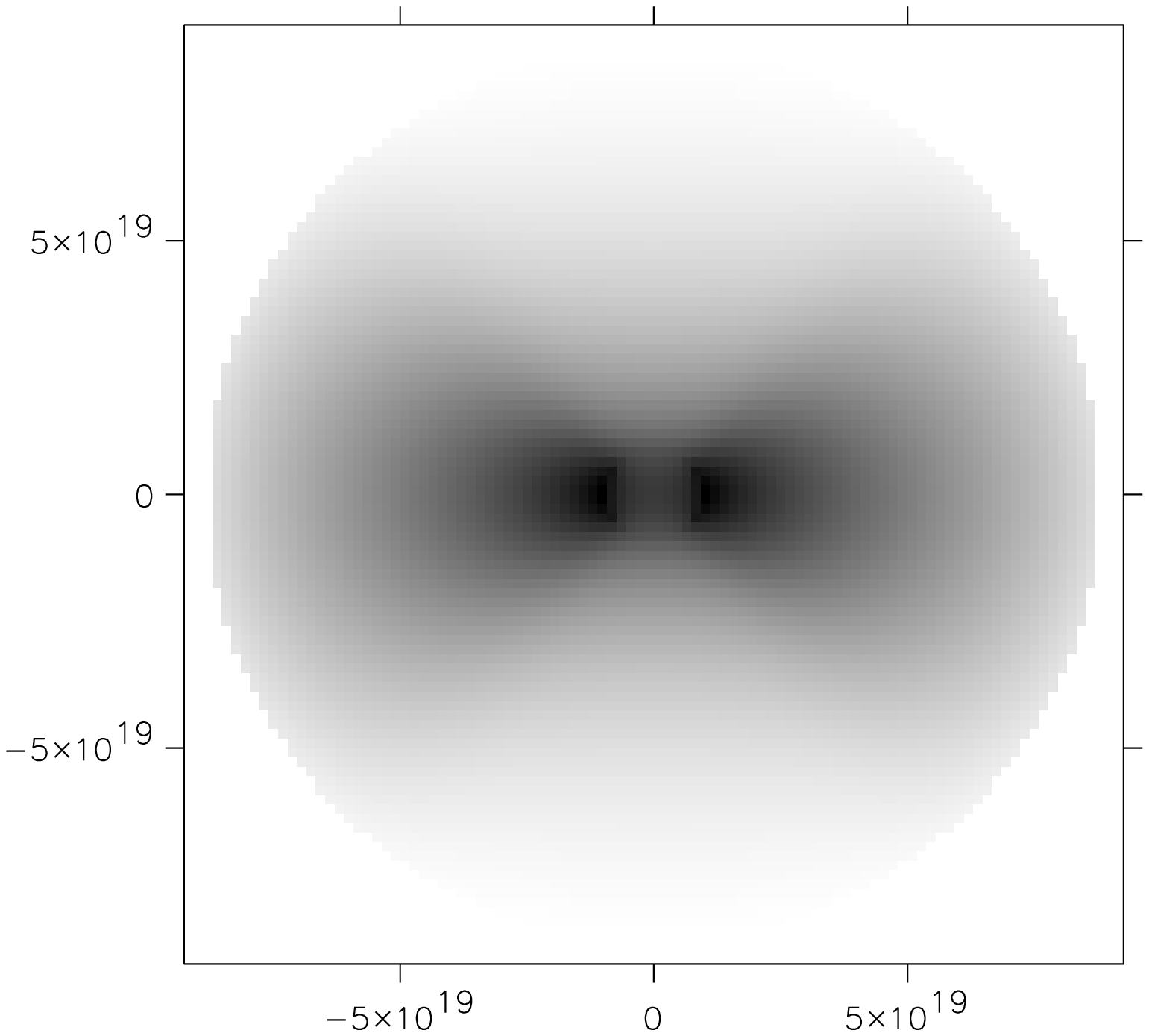}}
\hfill

\vspace{-1mm}
\caption{\label{7_images}
Images of the reference model, using a circular aperture. 
Top panels are for 30$^{\circ}$
inclination, bottom panels for 70$^{\circ}$. Left panels are
the 30\um\ images, right panels are at 200\um. The grey scale
is inverted: dark areas represent more emission. The central
white region is where the disk ends and the central emission
source is located. The outer circle represents the aperture.
}
\vspace{-3mm}
\end{center}
\end{figure*}

It is illuminating to examine the spatial distribution of
the emission in the disk. In Fig.~\ref{7_images} we show
images of the reference model, at 30 and 200\um,
and with inclination of 30 and 70$^{\circ}$. The images
display the entire angular grid, causing them to have a 
circular cut-off, independent of the inclination of the
disk. At 30\um\ the
disk is optically thick and only the inner regions 
radiate. At 70$^{\circ}$ the inner regions are not
seen directly, but some emission is seen from the
opposite inner side of the torus. At 200\um\ the dust
is optically thin and the entire disk will radiate.
At 30$^{\circ}$ the outermost regions are not as bright 
as the rest, since they are somewhat cooler. At 70$^{\circ}$
the inner regions are now seen direct, since the dust
is transparent. They are the brightest part of the
disk, since this is where the column density is
highest.

\begin{figure*}[t]
\begin{center}
\resizebox{3.3cm}{!}{\includegraphics{37611733.ps}}
\resizebox{3.3cm}{!}{\includegraphics{37611732.ps}}
\resizebox{3.3cm}{!}{\includegraphics{37611773.ps}}
\resizebox{3.3cm}{!}{\includegraphics{37611772.ps}}
\\
\resizebox{3.3cm}{!}{\includegraphics{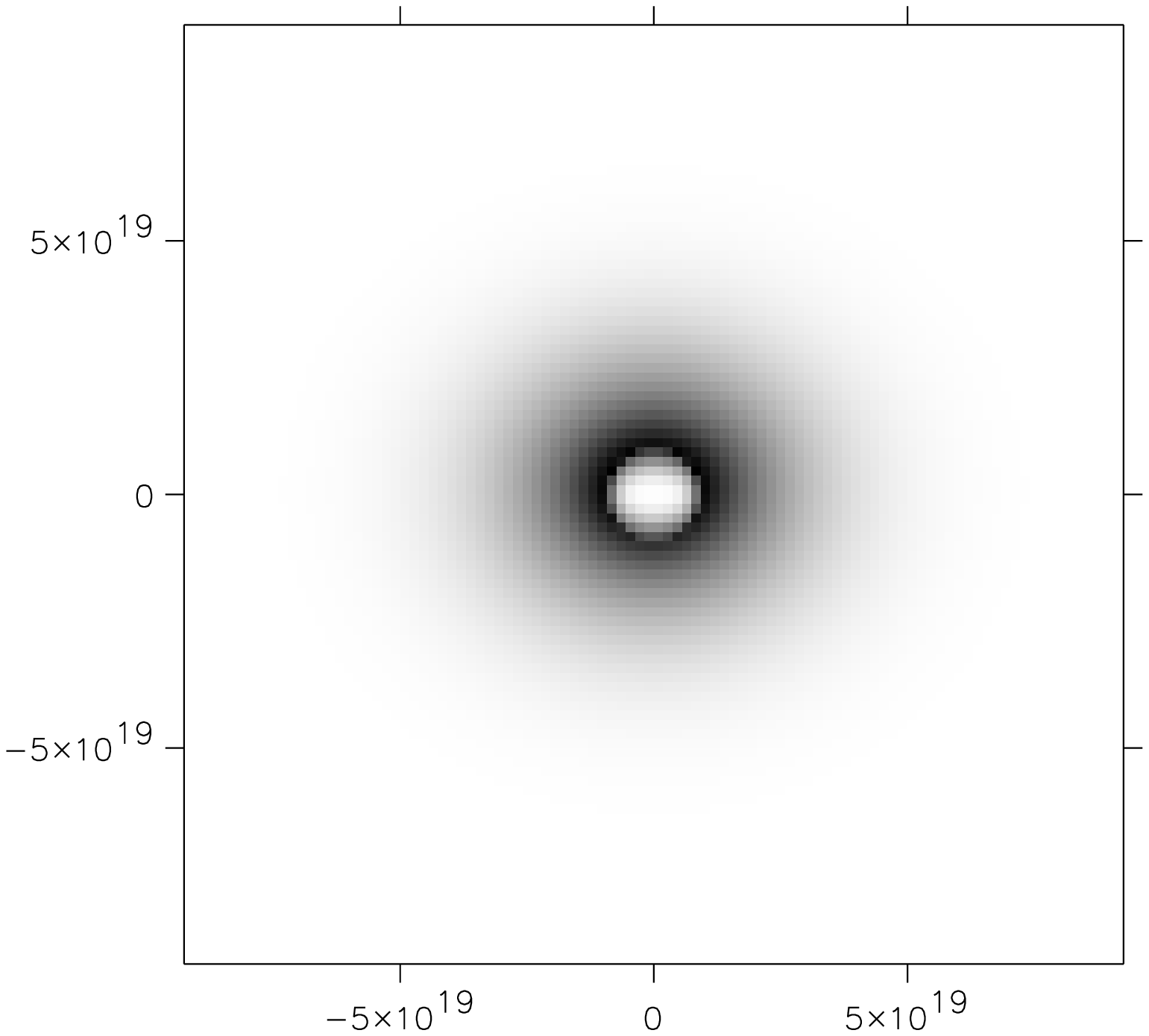}}
\resizebox{3.3cm}{!}{\includegraphics{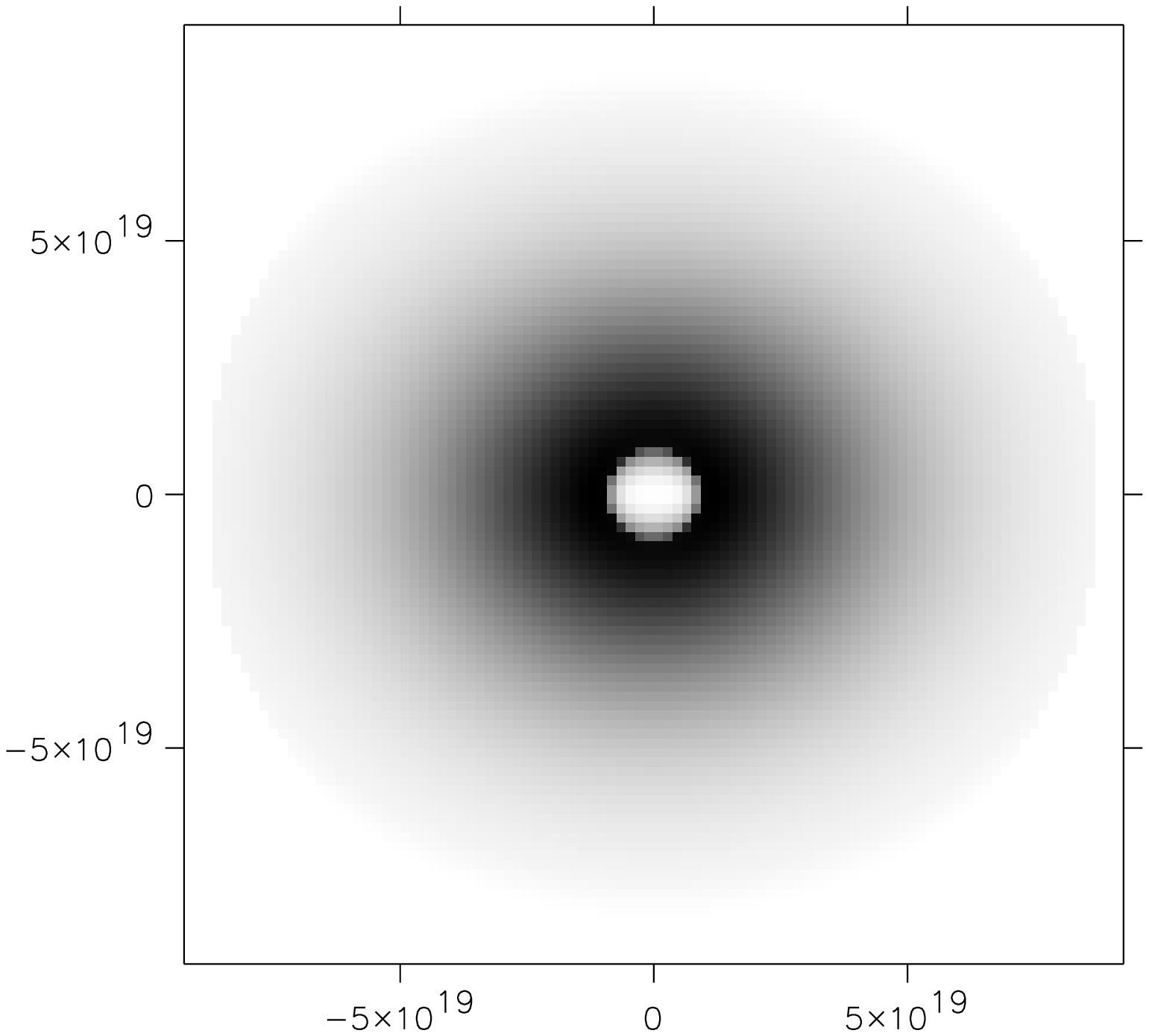}}
\resizebox{3.3cm}{!}{\includegraphics{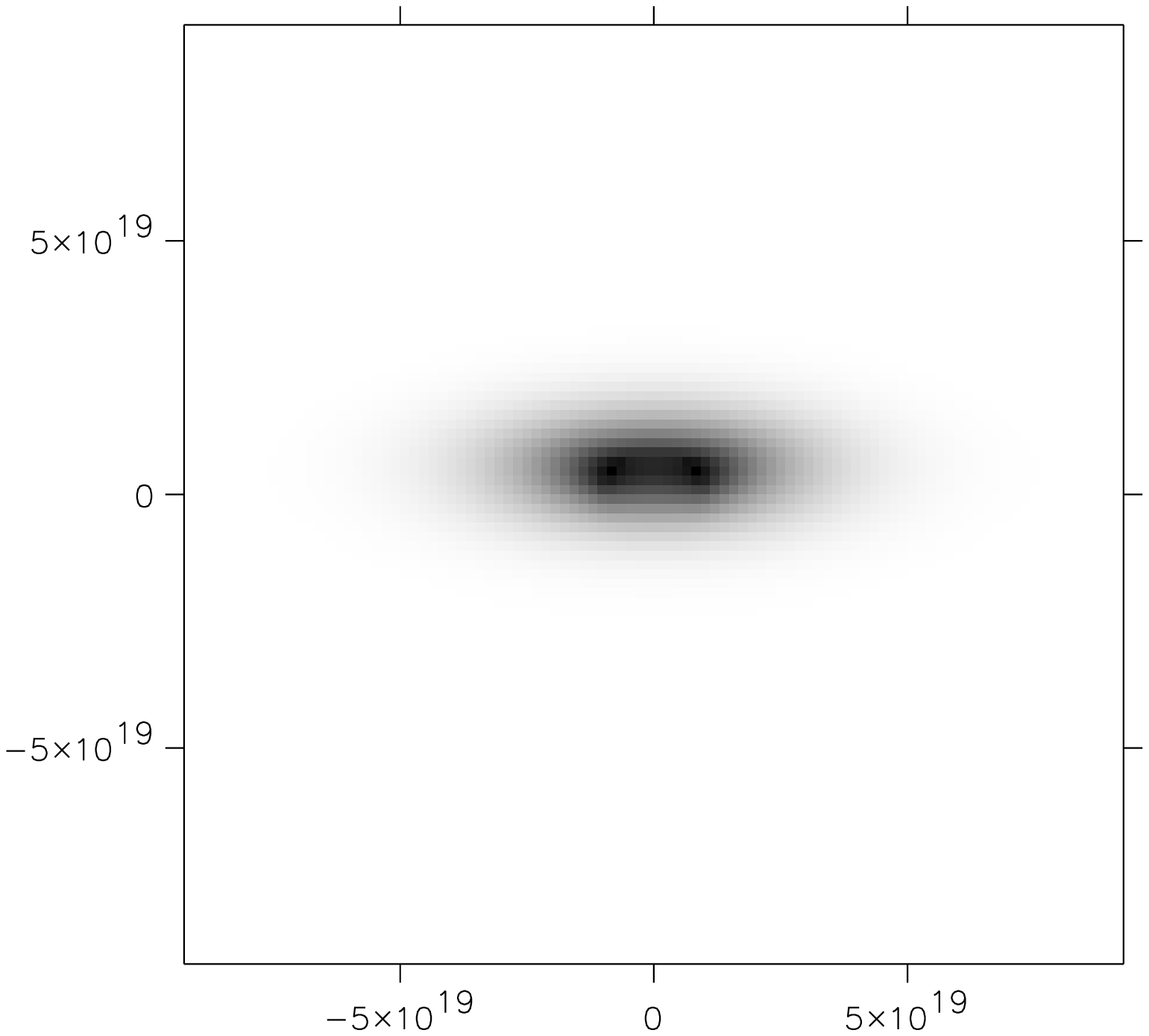}}
\resizebox{3.3cm}{!}{\includegraphics{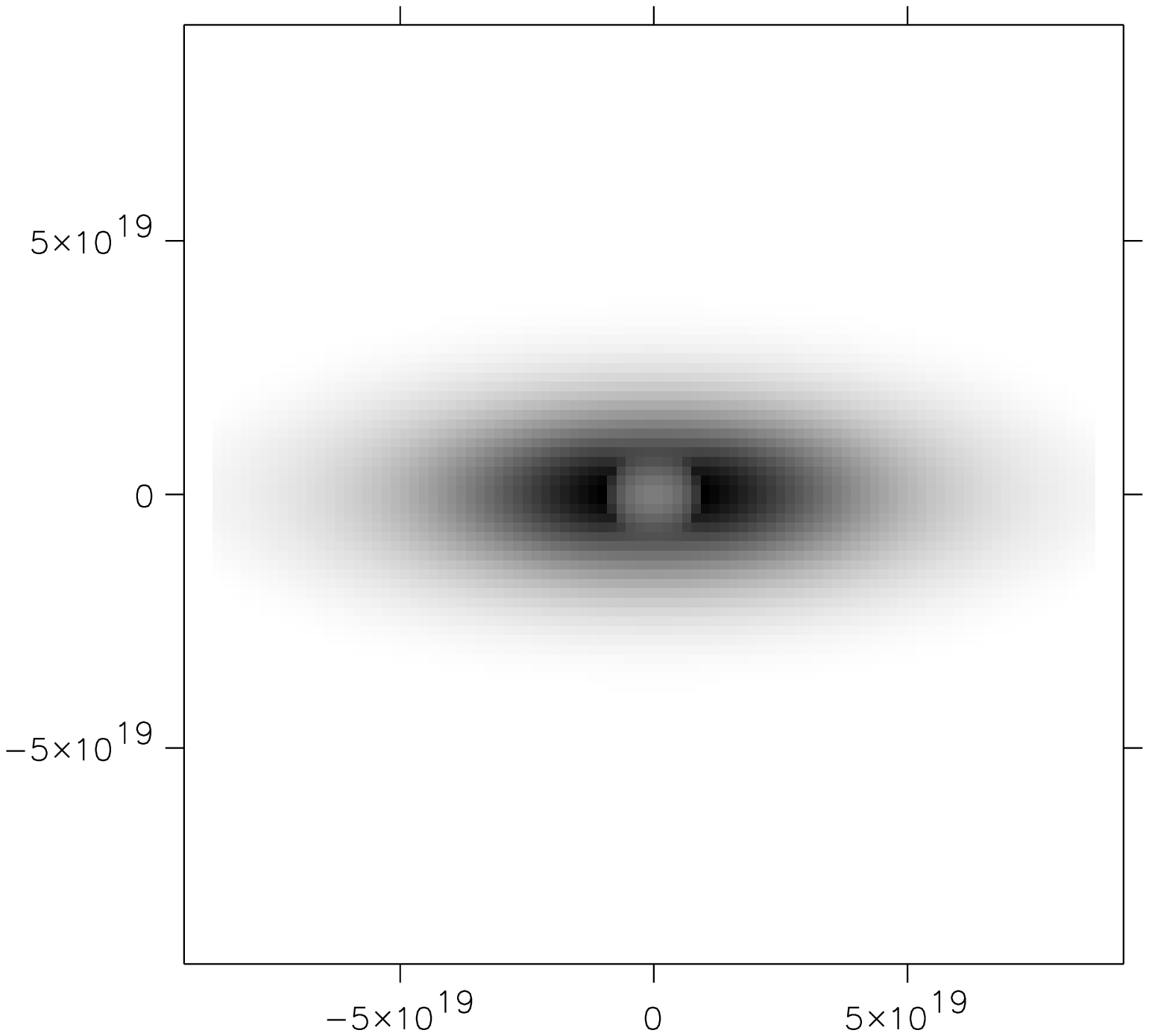}}
\\
\resizebox{3.3cm}{!}{\includegraphics{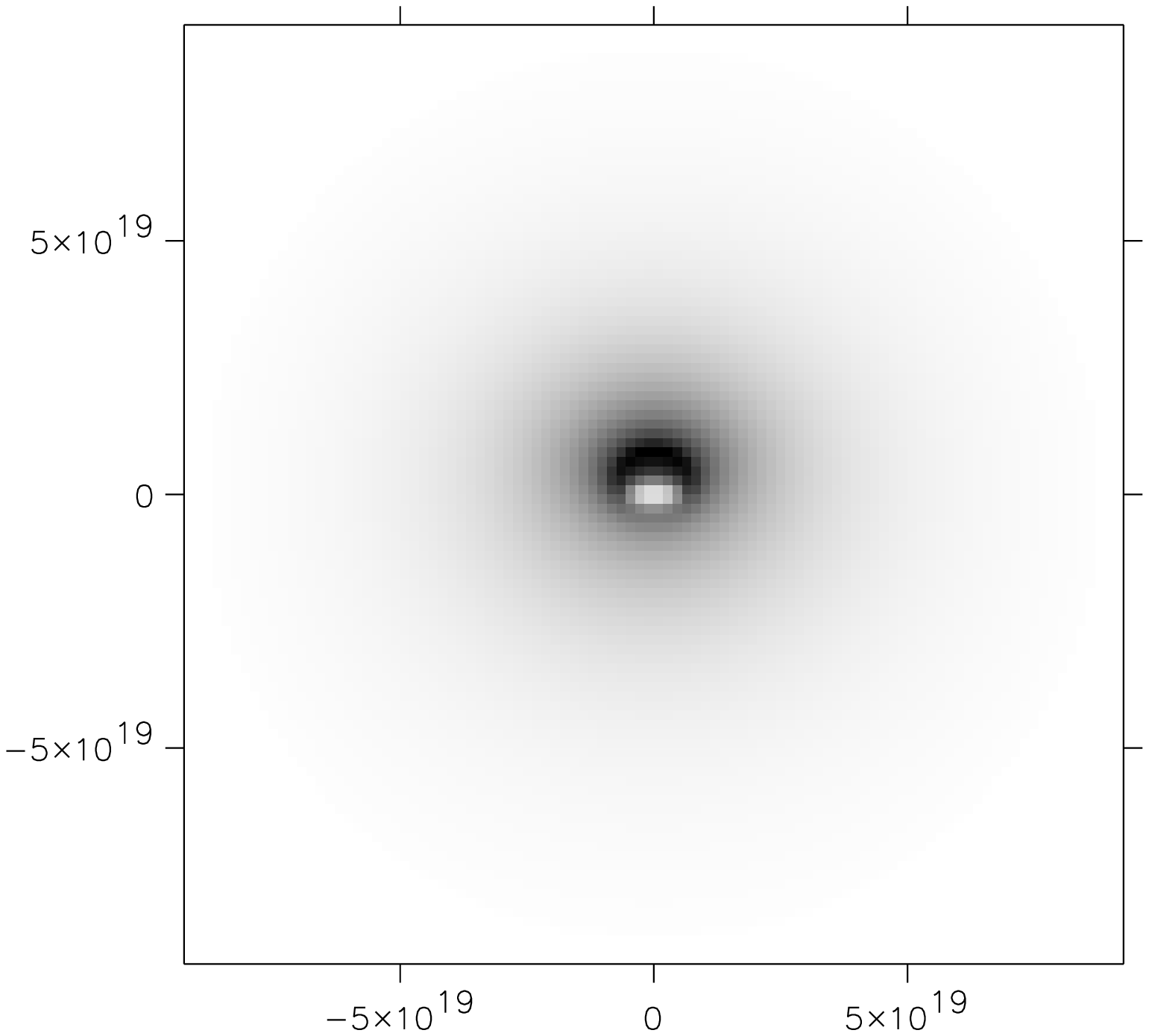}}
\resizebox{3.3cm}{!}{\includegraphics{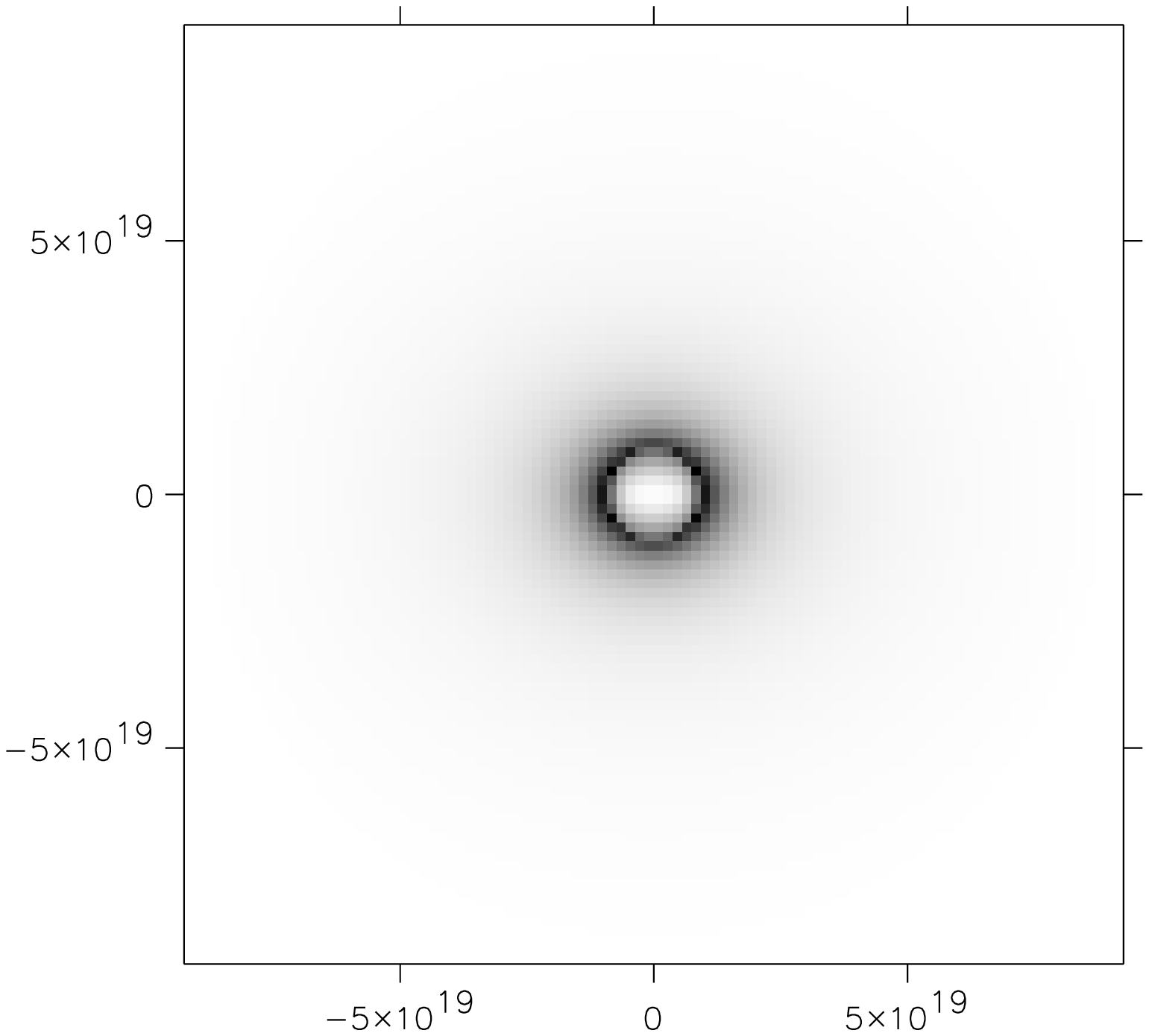}}
\resizebox{3.3cm}{!}{\includegraphics{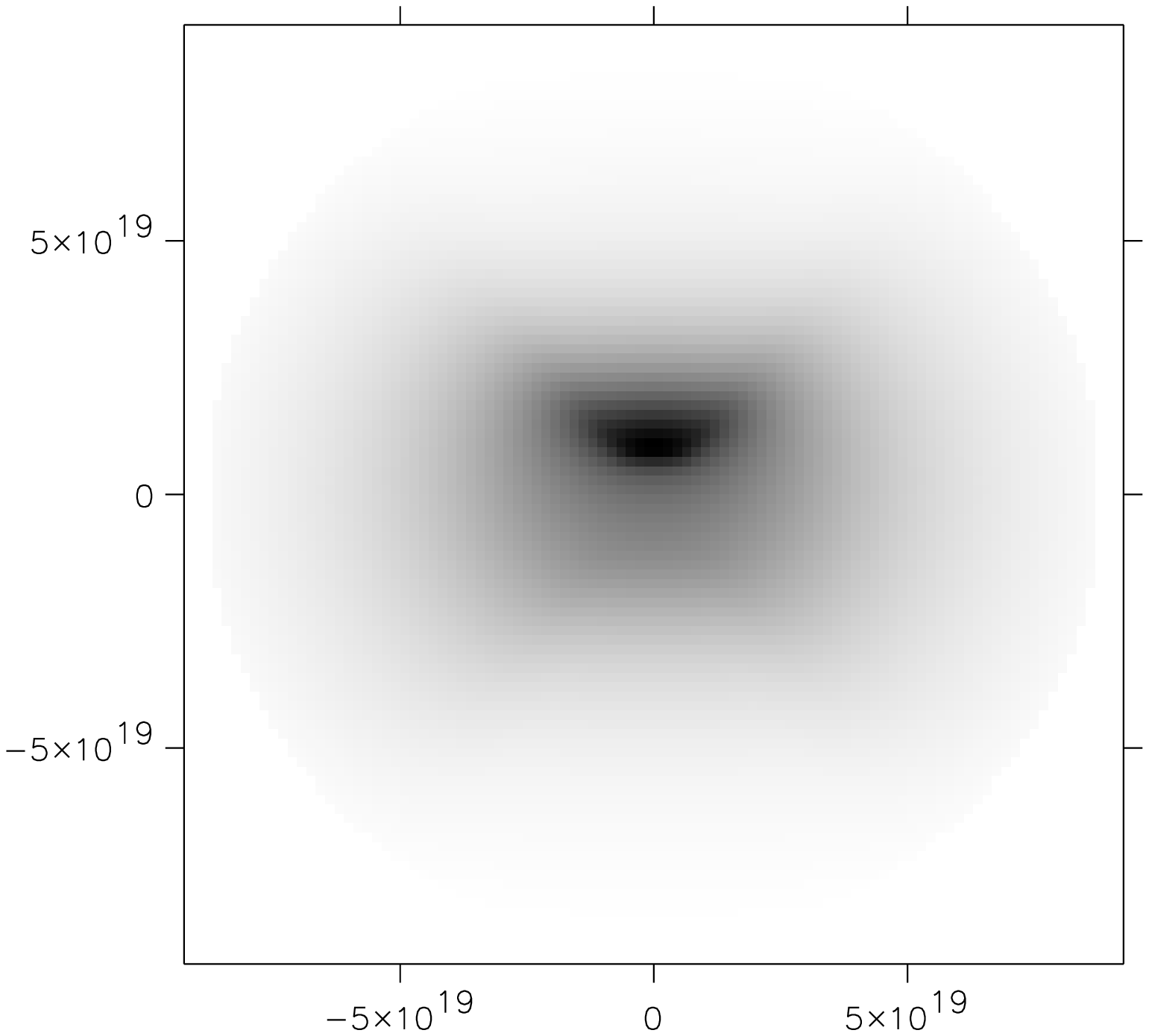}}
\resizebox{3.3cm}{!}{\includegraphics{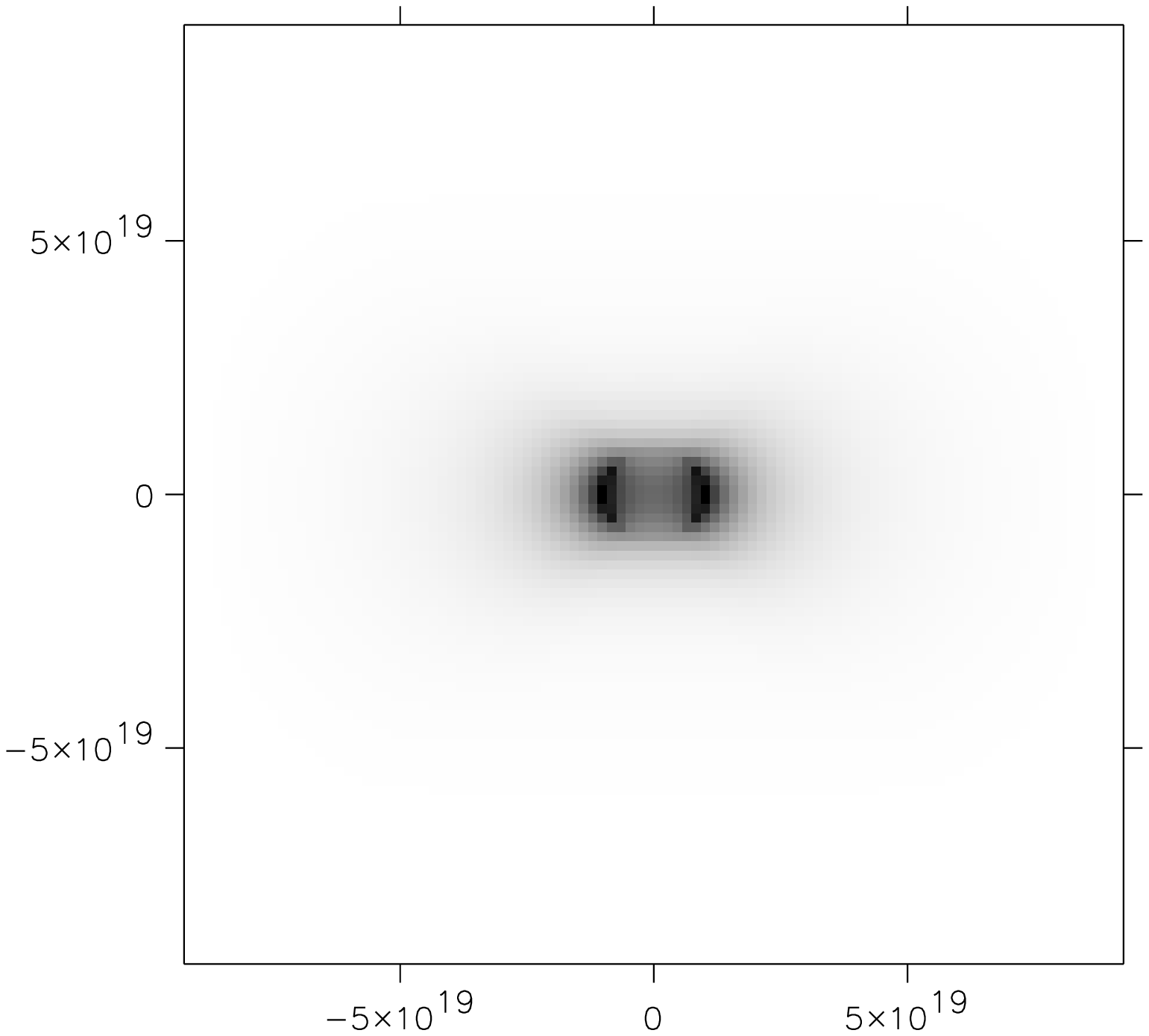}}
\\
\resizebox{3.3cm}{!}{\includegraphics{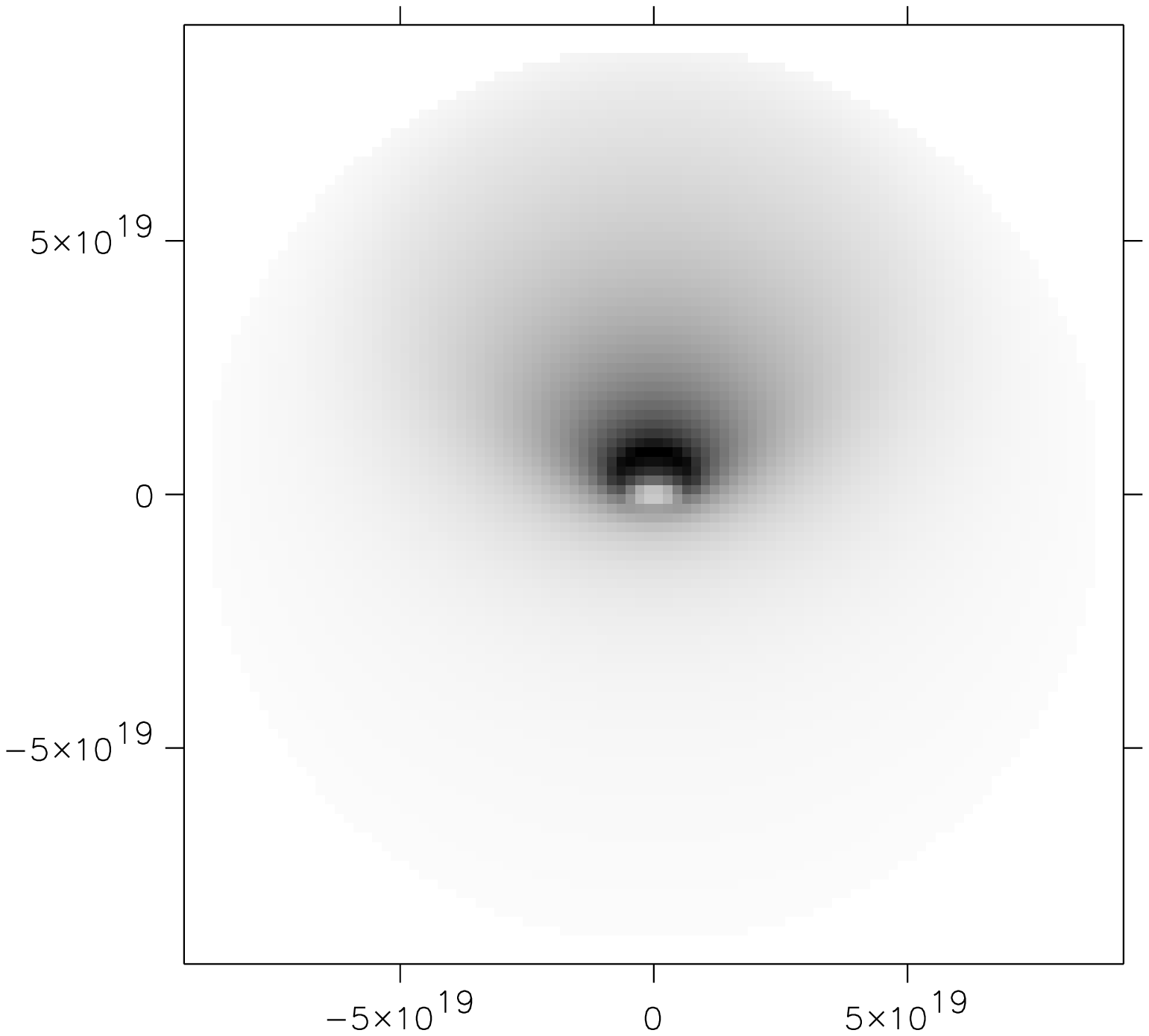}}
\resizebox{3.3cm}{!}{\includegraphics{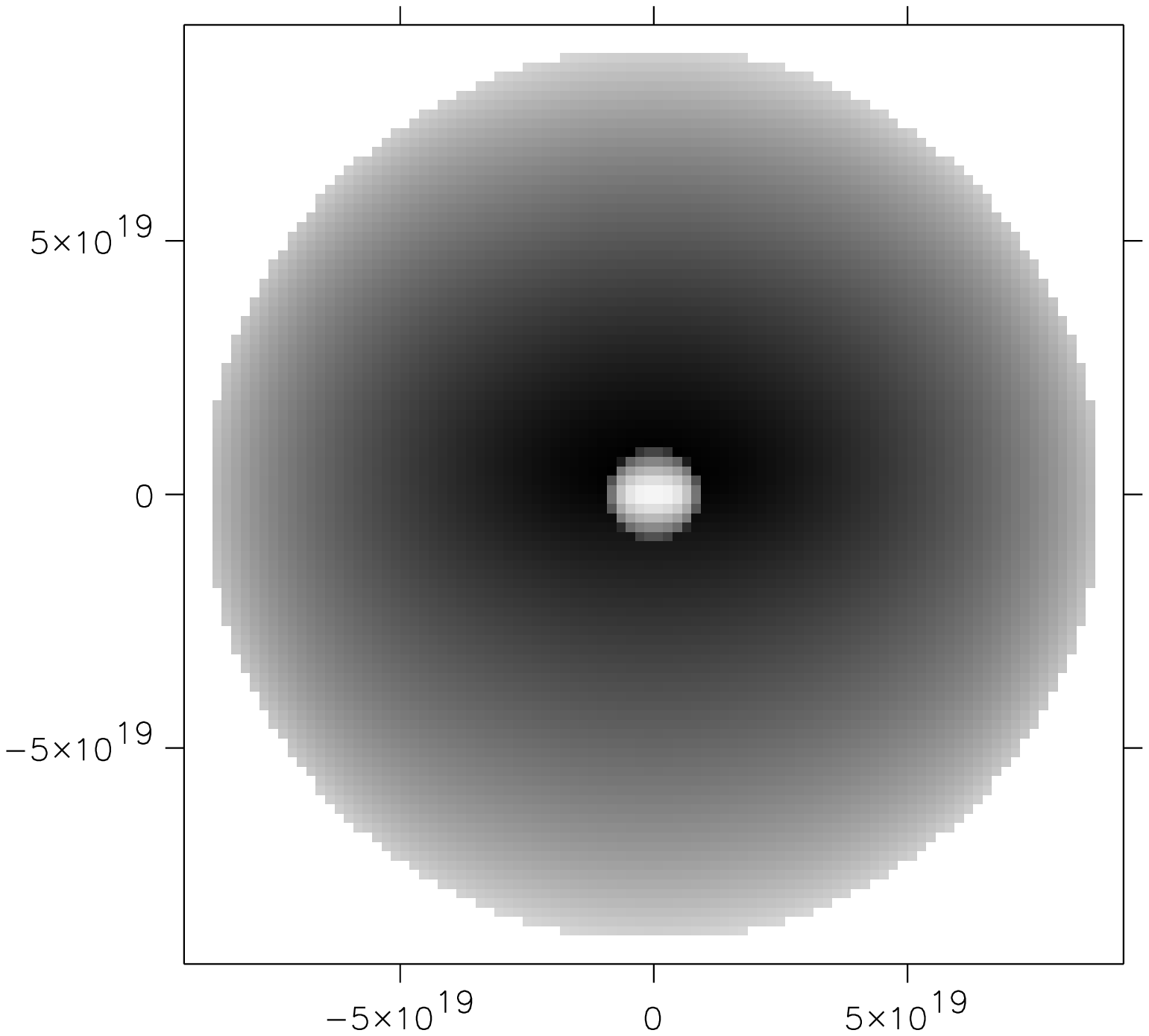}}
\resizebox{3.3cm}{!}{\includegraphics{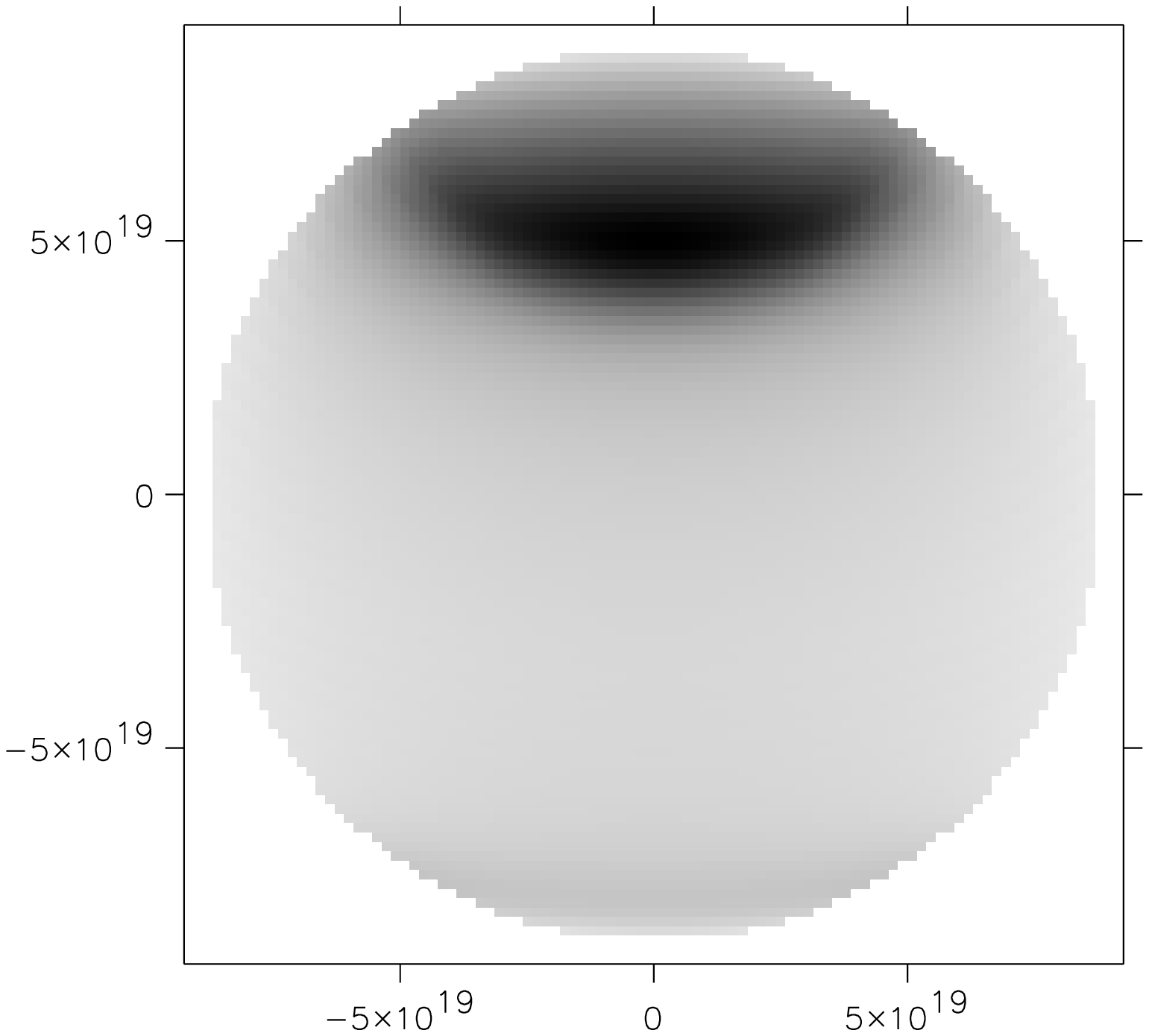}}
\resizebox{3.3cm}{!}{\includegraphics{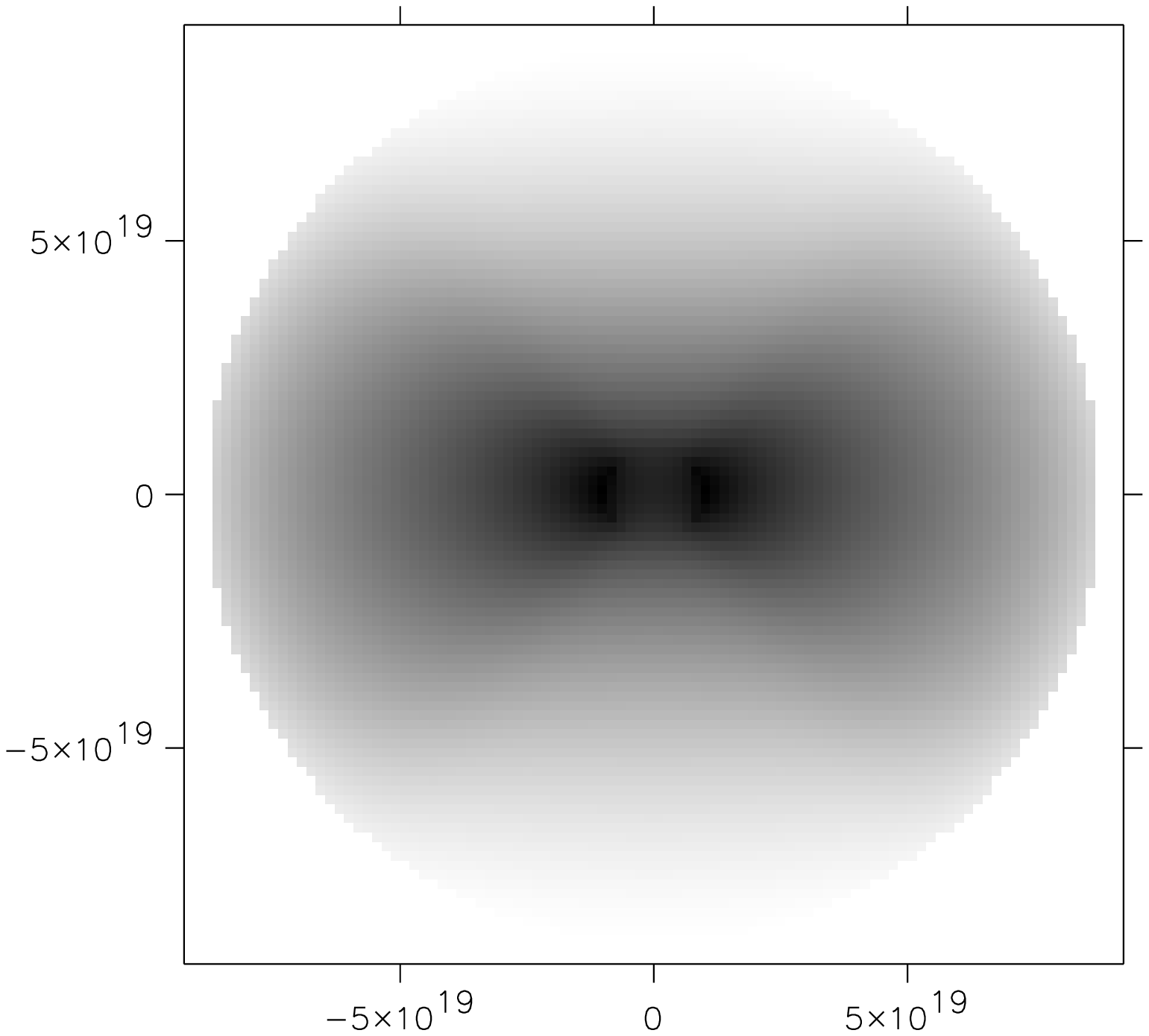}}

\vspace{-1mm}
\caption{\label{7_im_vars}
Images of the disk when varying the different parameters.
The top row is the reference model, then
from top to bottom per row are changed: $\alpha=0$, 
$\beta=-2$ and $\tau_{20}=50$. From left to right the
images are for 30$^{\circ}$ at 30 and 200\um, and
70$^{\circ}$ at 30 and 200\um\ respectively.
}
\end{center}
\vspace{-3mm}
\end{figure*}

The effect of varying the parameters is shown in 
Fig.~\ref{7_im_vars}. The top row shows again the
reference model. The second row shows the pillbox
models ($\alpha=0$), which are clearly less bright.
This difference is most pronounced at 200\um. They 
are of course flatter at 70$^{\circ}$. The
third row displays the $\beta=-2$ models, i.e. models
with equal mass per logarithmic interval. Here the
inner regions dominate all the images, since they
represent the densest parts of the disk. This is especially
clear in the 200\um\ images, where the disk is
optically thin. The final row contains the images
for a disk with $\tau_{20}=50$. At 30\um\ the emission 
is more centrally concentrated and the contrast with
the outer regions is larger than for the reference
model. However, for the 200\um\ images the opacity
is irrelevant, and the images are comparable. 

\subsection{Comparison with models in literature}
The resulting spectrum from our reference model
can be compared to literature models with comparable disk and 
dust parameters. The main difference between our model and the 
literature models is the absence of significant 10\um\ emission 
at small inclinations, and only weak 10\um\ absorption at larger 
inclinations. However, since all literature models use Standard 
Galactic Dust, a comparison at 10\um\ is unrealistic.

A qualitative comparison of our reference model with literature 
models with comparable parameters reveils that the general 
shape of our output spectrum is best matched by the \citet{pk92} 
models (hereafter PK92). The models by \citet{gra94} (hereafter
GD94) also match the shape quite 
well, but the \citet{efst95} (hereafter ER95) models are very 
different, with a
much flatter SED in the mid-infrared and often a peak in the
near-infrared. Our reference model becomes optically thin longward 
of 100\um. Only the PK92 models match this, other models 
all become optically thin between 60 and 80\um. 
In all literature models changing the inclination causes
a continous decrease in flux at a fixed wavelength. The
GD94 models show a very discrete change in flux between
45 and 75 degrees inlination. As a consequence, there is no 
obvious difference between their models with 0 and 45 degrees 
inclination, and neither between the models with inclinations
of 75 and 90 degrees, except for the very thick and compact
geometries.

For a more quatitative comparison we have calculated an anisotropy 
value, 
defined by taking the flux ratio for face-on over edge-on oriented 
disks at 20\um. For the literature models that have parameters 
closest to our reference model, we find this ratio to be $\sim 5$. 
This is consistent with the anisotropy value in our reference 
model. 

Clumpy torus models, as described by \citet{nenk02} are in
general flatter than our reference model, which is especially
prominent at small inclinations. Their anisotropy value is
also $\sim5$, as in our reference model, and they do not
become optically thin shortward of 100\um. Depending on
the mean free path of the clumps, the models have a more
pronounced peak around 60\um, and become more luminous.

A more detailed comparison is not feasible, since the models
in literature use different setups and calculation methods. All 
observed differences between our results and literature models
can be ascribed to this.

\section{Understanding the behaviour}
The output of a dusty disk depends mainly on two properties: the
temperature and the mass of the dust. If the dust is further
away from the emission source, it will be cooler, and therefore
emit at longer wavelengths. At shorter
wavelengths the dust is self-shielding, since the dust is optically
thick for its own radiation. The
inner regions will become very hot, but these can only get
rid of the energy at wavelengths where the dust is optically
thin. As a consequence, the inner regions will be the brightest at the
shorter wavelengths, and only in the far-infrared will the
entire disk start to radiate. This effect is 
clearly seen in the images (Fig.~\ref{7_images}).

\begin{figure*}[!Ht]
\begin{center}
\resizebox{16.cm}{!}{\includegraphics{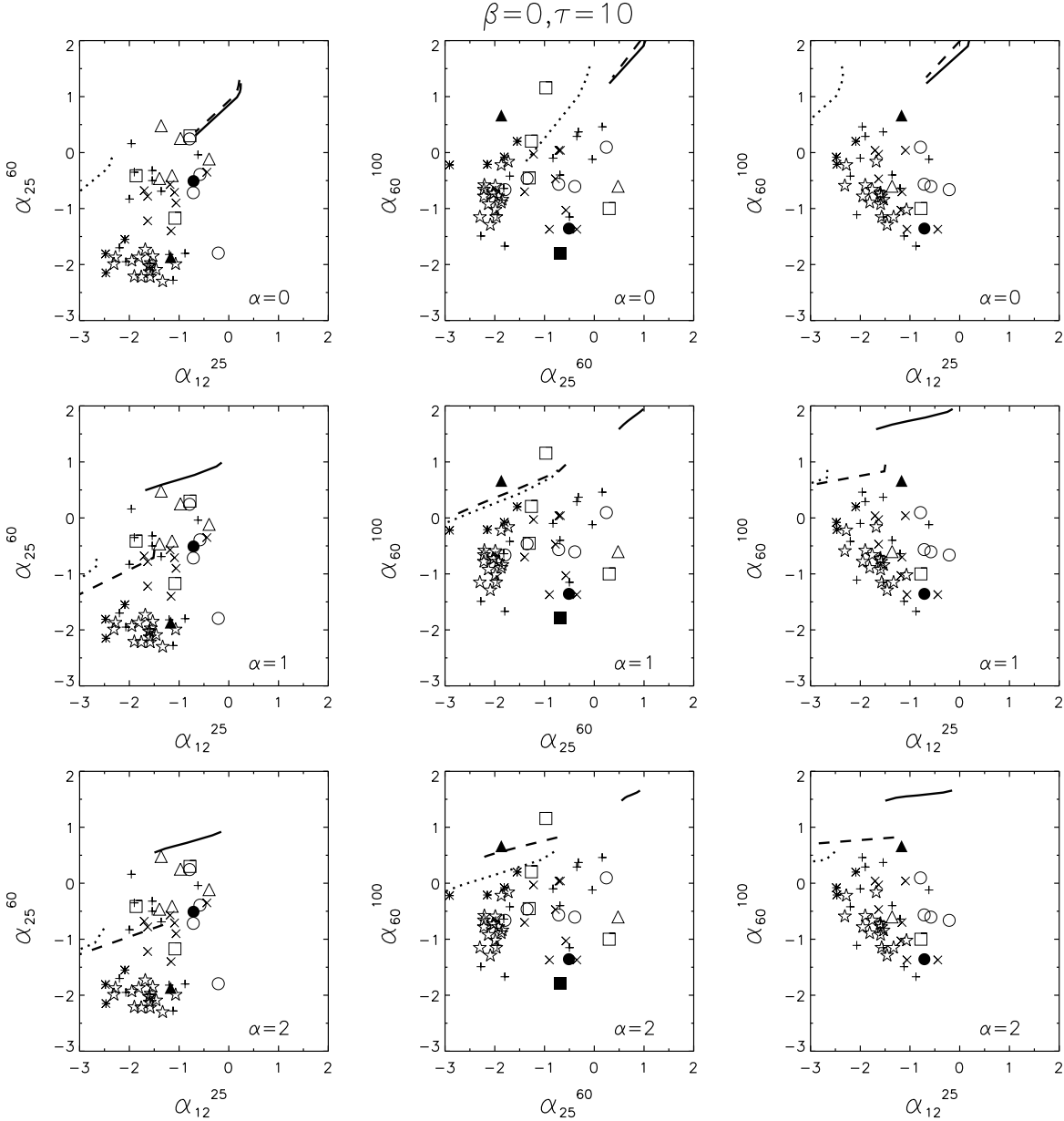}}
\hfill
\vspace{-1mm}
\caption{\label{7_cc_alfa}
Colour-colour plots for models with $\beta=0$, $\tau_{20}=10$ and varying
$\alpha$ and disk size. Lines indicate varying inclination: from left
to right corresponds to edge-on to face-on orientations.
Solid lines are for the
inner disk, dotted for outer disk, dashed for large disk. 
The objects are coded as follows: open circles=quasars, open 
squares=narrow-line radio galaxies, open triangles=broad-line radio 
galaxies, filled circle=Heckman quasar sample, filled
square=Heckman radio galaxy sample, filled triangle=\cyga, 
crosses=Seyfert~1, plusses=Seyfert~2, asterisks=ultra-luminous galaxies, 
open stars=starburst galaxies.
}
\end{center}
\vspace{-3mm}
\end{figure*}

\begin{figure*}[!Ht]
\begin{center}
\resizebox{16.cm}{!}{\includegraphics{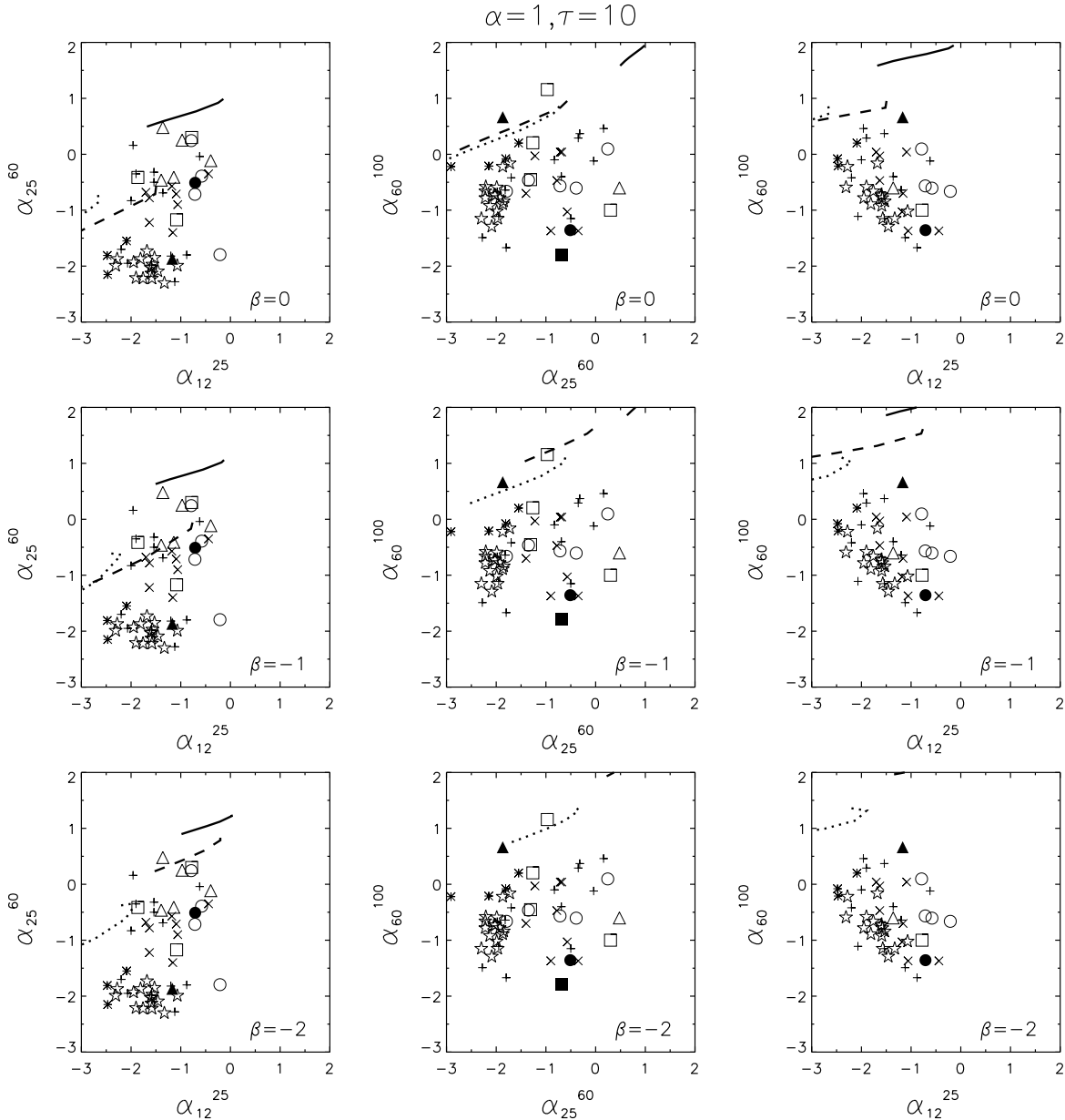}}
\hfill
\vspace{-1mm}
\caption{\label{7_cc_beta}
Colour-colour plots for models with $\alpha=1$, $\tau_{20}=10$ and varying
$\beta$ and the disk size. Lines and symbols as in Fig.~\ref{7_cc_alfa}.
}
\end{center}
\vspace{-3mm}
\end{figure*}

\begin{figure*}[!Ht]
\begin{center}
\resizebox{16.cm}{!}{\includegraphics{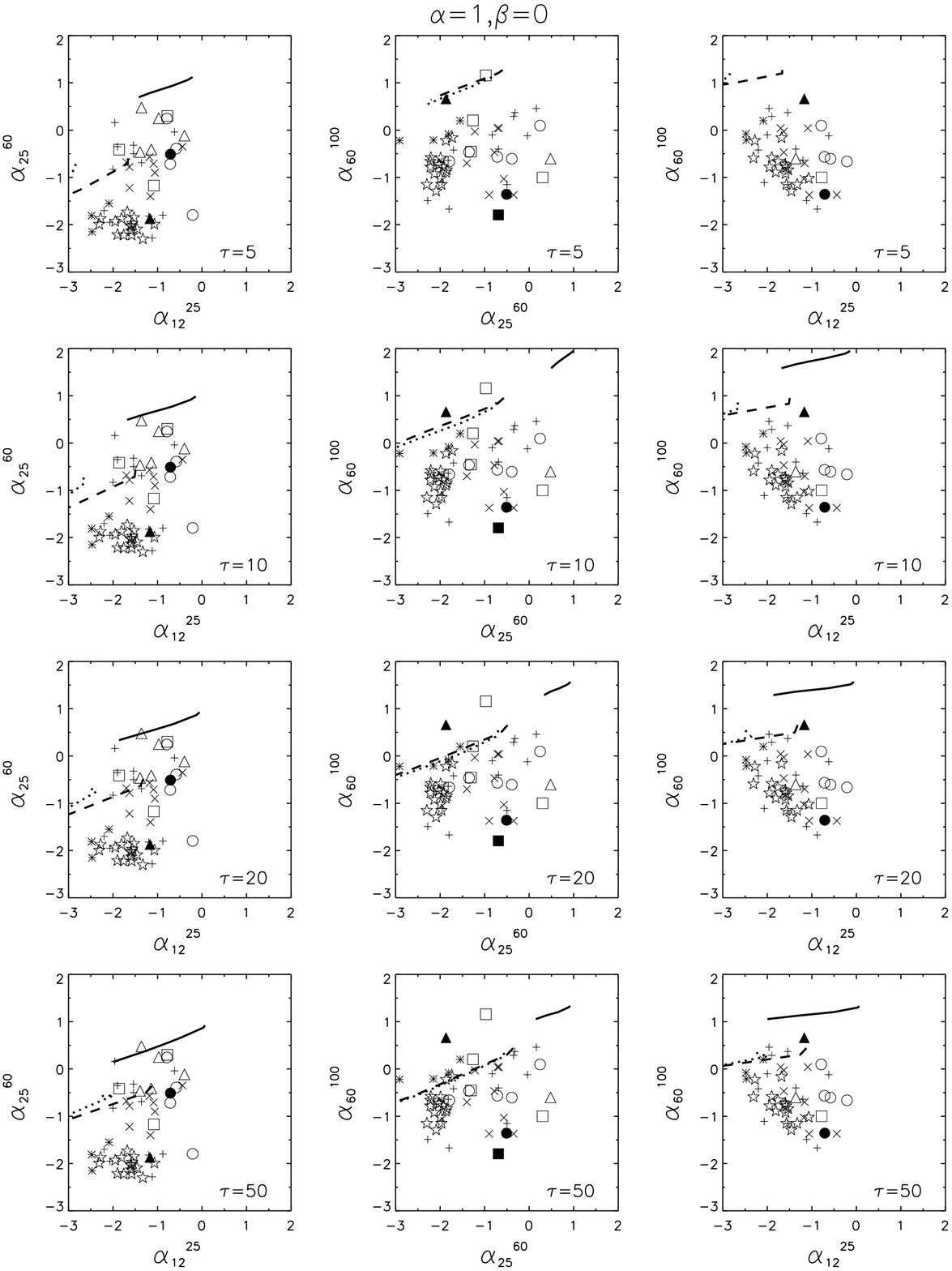}}
\hfill
\vspace{-1mm}
\caption{\label{7_cc_tau}
Colour-colour plots for models with $\alpha=1$, $\beta=0$ and varying
$\tau_{20}$ and disk size. Lines and symbols as in Fig.~\ref{7_cc_alfa}.
}
\end{center}
\vspace{-3mm}
\end{figure*}

In a pillbox model, the $\tau=1$ surface is 'squeezed' more to
the equatorial plane and, hence, compared to the other models less
radiation of the central source is intercepted. Pillbox models are
therefore less luminous. The difference between conical and flaring
disks is less pronounced. Nevertheless, for a flaring disk, some 
dust at large radii will have a direct line of sight to the 
central source and be warmer. This effect is best observed at 
large inclinations (see Fig.~\ref{7_model_compare70}).

When we vary $\beta$ from 0 to --2, more mass will be
concentrated in the inner regions. This dust will be hotter,
and therefore brighter in the near- and mid-infrared. Since
the dust is more centrally concentrated for $\beta=-2$, there 
is less mass in the outer regions, and therefore less cool 
dust, as seen in Fig.~\ref{7_im_vars}. As a
consequence the far-infrared emission will decrease in flux, and
the entire spectrum shifts to shorter wavelengths.

Increasing the optical depth increases the amount of absorbed
radiation. For higher values of $\tau_{20}$ it will be more 
difficult to transport
absorbed energy into the torus. As a consequence the inner parts 
get hotter, but the outer parts will remain cooler. This results in 
a wider temperature distribution, which translates into a broader
energy distribution for small inclinations (Fig.~\ref{7_model_compare30}).
For large inclinations, the increased amount of cold dust is most
important. Since the inner regions are shielded from direct
view, the energy distribution shifts to longer wavelengths.

The distance to the central source is critical in the
models. At large inclinations, we cannot view the inner edge of 
the torus direct. The hot dust is shielded from view by the cool
outer regions. For this reason, the large disk and outer disk 
have comparable spectra. Overall the inner disk is much warmer
and brighter at the shortest wavelengths. At small inclination,
the whole disk is visible and both the inner and large disk,
which have hot dust close to the central source, are much brighter
at shorter wavelengths. The large disk and the outer disk show very
similar behaviour at the longest wavelengths where the dust is 
optically thin. The large disk produces less near- and
mid-infrared emission than the inner disk, even though its
inner radius is similar, since the temperature gradient is
steeper. This is also related to the model setup: since the
disk is larger, a smaller value for $\Sigma_0$ is required
to obtain the requested equatorial optical depth, and hence
the total opacity is lower.

Assuming $q=-2$ instead of using Galactic dust mainly affects
the 10\um\ silicate feature.
Although our models clearly lack significant silicate emission,
they still display clear absorption, albeit not as deep as the
Galactic dust models. The detailed behaviour of the silicate
feature will be the purpose of future high frequency resolution
models.

\section{Comparing the models to observations}
\subsection{Colour comparison}
Using the known infrared colours of samples of active galaxies, we can
attempt to constrain the parameters of the obscuring medium. We
will use the IRAS bands (12, 25, 60 and 100\um) which are also
available as ISO bands. We have selected all 3C radio-loud
active galaxies from the literature \citep{gol88,neu86}
with a least three solid
detections in three of the IRAS bands, and with small redshifts
($z < 0.5 $). It should be mentioned, however, that IRAS detected
only a few radio-loud active galaxies, which all lie close to
the detection limits of the instrument. In addition we selected
a sample of starburst, ultra-luminous and
Seyfert galaxies with four IRAS detections from literature
\citep{sek87,mack89}.

For all radio-loud objects we assured that the infrared 
emission is of thermal nature. From \citet{heck92}
we have constructed the colours for samples of quasars
and radio galaxies. All data are overplotted on the models,
while varying different parameters in Fig.~\ref{7_cc_alfa}, \ref{7_cc_beta}
and \ref{7_cc_tau}. The $\alpha$ symbol on the axis, annotated
with two wavelengths, represents
the spectral index or colour, defined as $F \propto \nu^{\alpha}$,
while the $\alpha$ symbol in the plots, without annotation,
refers to the geometry of the disk. We will refer to the
colours as hot ($\alpha_{12}^{25}$), warm ($\alpha_{25}^{60}$),
and cold ($\alpha_{60}^{100}$). For an increase in spectral
index the flux at the shorter wavelength increases with respect to
the flux at the longer wavelength. Hence an increase in the value
for the colour implies more flux at shorter wavelength.

The objects within a given class cluster together. The quasars
have a narrow range of hot and cold colours, both lie between
0 and $-1$. The warm colour ranges from 0.3 to $-2$. The
narrow-line radio galaxies have cooler hot colours than 
the quasars, ranging from $-0.5$ to $-2$. The warm colours show
approximately the same values, from 0 to $-1.5$, but the cold
colour has a large range from 1 to almost $-2$. For the broad-line
radio galaxies the hot colour is comparable to that of
the quasars, but the warm colour is clearly warmer, having
typical values between 1 and $-0.5$. The cold colour is only
determined for one object, the rest lack 100\um\ detections.
The starburst and ultra-luminous galaxies lie clearly in a
different region than the active galaxies
in the hot vs. warm plot, while their cold colours are comparable to 
the cold colours of the radio-loud active galaxies. The Seyfert~1
colour distribution is indistinguishable from the colours for
the radio-loud active galaxies, but the Seyfert~2 galaxies overlap
both the active galaxies and the starbursts in their hot and warm
colours. 

There are no objects which have a positive hot colour in the entire
sample,
and only a few which have a positive warm colour. Nevertheless, the 
colours of the three radio-loud classes do not separate out too much. 
From an infrared point of view, these three classes are a somewhat 
arbitrary classification of a continuum distribution of types.

\subsubsection{Models with $\beta=0$ and $\tau_{20}=10$}
In Fig.~\ref{7_cc_alfa}, we have fixed $\beta$ and $\tau_{20}$, and
varied the disk size (3 different lines in plot), $\alpha$
(the different plots) and the orientation (along the lines,
edge-on to face-on is from left to right). The overplotted data
are for objects which are presumed to represent a continuum
of different orientations, from quasars to narrow-line radio 
galaxies. The filled symbols are for ensembles
of quasars and radio galaxies.\\
Changing $\alpha$ from $\alpha=0$ (pillbox) to $\alpha=1$ (conical) has
a profound effect on the colours of the disk. A further increase to
$\alpha=2$ (flaring disk) has little effect. These effects are in 
line with the SEDs shown in Figs.~\ref{7_model_compare30} and
\ref{7_model_compare70}.

\subsubsection{Models with $\alpha=1$ and $\tau_{20}=10$}
In Fig.~\ref{7_cc_beta}, the value for $\alpha$ is fixed, as
is $\tau_{20}=10$. Decreasing $\beta$ to
$-2$ the models move up, except the outer disk, indicating 
more positive warm and cold spectral indices. The hot colour
only increases with decreasing $\beta$ in the large disk.

\subsubsection{Models with $\alpha=1$ and $\beta=0$}
So far we have fixed the value of $\tau_{20}$, but this can of course
also vary. In Fig.~\ref{7_cc_tau} we have fixed $\alpha=1$ and
$\beta=0$ and varied $\tau_{20}$ in the different plots. Along
the lines we still vary the inclination.\\
Increasing $\tau_{20}$ will widen the range of the hot and
warm colours of the models with changing the inclination. The
warm colour also becomes cooler with increasing opacity. The 
cold colours have approximately the same range when varying
$\tau_{20}$, but become cooler for higher values.

\subsubsection{Comparison with observations}
As judged from the plots, for all models the cold colours 
appear inconsistent with the observations. This implies that 
either the observed 60\um\ flux or the 100\um\ flux, or both are 
not fitted by the models. The hot colours in the models span
the same range as the observed colours for active galaxies, but 
in the hot vs warm colour plot, the warm 
colours in the models do not cover the entire range of observed values. 
This indicates that already at 60\um\ the spectrum starts to 
deviate. Hence, both the 60 and 100\um\ emission in the active
galaxies are dominated by a different component. 

At shorter wavelengths, the difference in hot and warm colours
between starbursts and active galaxies indicates that the 12 and
25\um\ emission in active galaxies arises from a component unrelated
to star-formation, most likely the active nucleus. Longward
of 60\um\ the colours of active galaxies are indistinguishable
from the colours of non-active galaxies, which implies that the
emission process is the same in both groups. This favours
star-formation as the dominant process of far-infrared emission
in active galaxies. A similar conclusion is reached by studies of 
infrared colours of Seyfert galaxies longward of 100\um\ by 
\citet{spi02}, who find that this emission is independent of the 
presence of an active nucleus. Furthermore, \citet{prie01} conclude
that coronal line strength is strongly correlated with near- and
mid-infrared emission, but not at all with far-infrared, indicative
of a different heating mechanism for the far-infrared emission.\\

In the hot vs warm colour plot the narrow-line and broad-line radio
galaxies are well fitted by the models. For the other colour plots
only a few narrow-line radio galaxies are still within the range of
model colours. The outlying quasar in the hot vs warm colours is
3C\,48, known to have a large star-formation rate, and possible all
its infrared emission arises in star-forming regions \citep{can00}.
Its warm colour is comparable to the warm colour of objects with known
large star-formation rates, but its hot colour is clearly warmer,
indicative of very hot dust.  This indicates that a significant
contribution from an active nucleus will cause objects to have a more
positive warm colour, while their cold colours are indistinguishable
from the colours of star-forming objects.

The inner disk does not fit the observed colours well, except
in the hot vs warm colour plot. The large and outer disk do better
for $\alpha=1$ and $\alpha=2$ disks and large inclinations, but
also fail in the hot vs cold colours. The pillbox disks do not
fit any of the observations. For decreasing $\beta$
all models move away from the observations. Increasing $\tau_{20}$
moves the models closer to the observed colours, but the
discrepancy in cold colours is not explained.\\

We note with interest that the ultra-luminous galaxies are
well fitted by the high opacity models in the cold vs. hot
and cold vs. warm colour plots. Also \cyga\ lies closer to
the models than the remainder of the radio galaxy population.
For the ultra-luminous galaxies it is known that star-formation
dominates the far-infrared output, and also \cyga\ is has strong
evidence for significant star-formation in the host galaxy. Hence,
star-formation can mimic the SED of an obscuring torus and
its relative contribution must be assessed by other means,
e.g. PAH emission, optical absorption lines, to
fully disentangle the two infrared components

\begin{figure*}[!t]
\begin{center}
\resizebox{8cm}{!}{\includegraphics{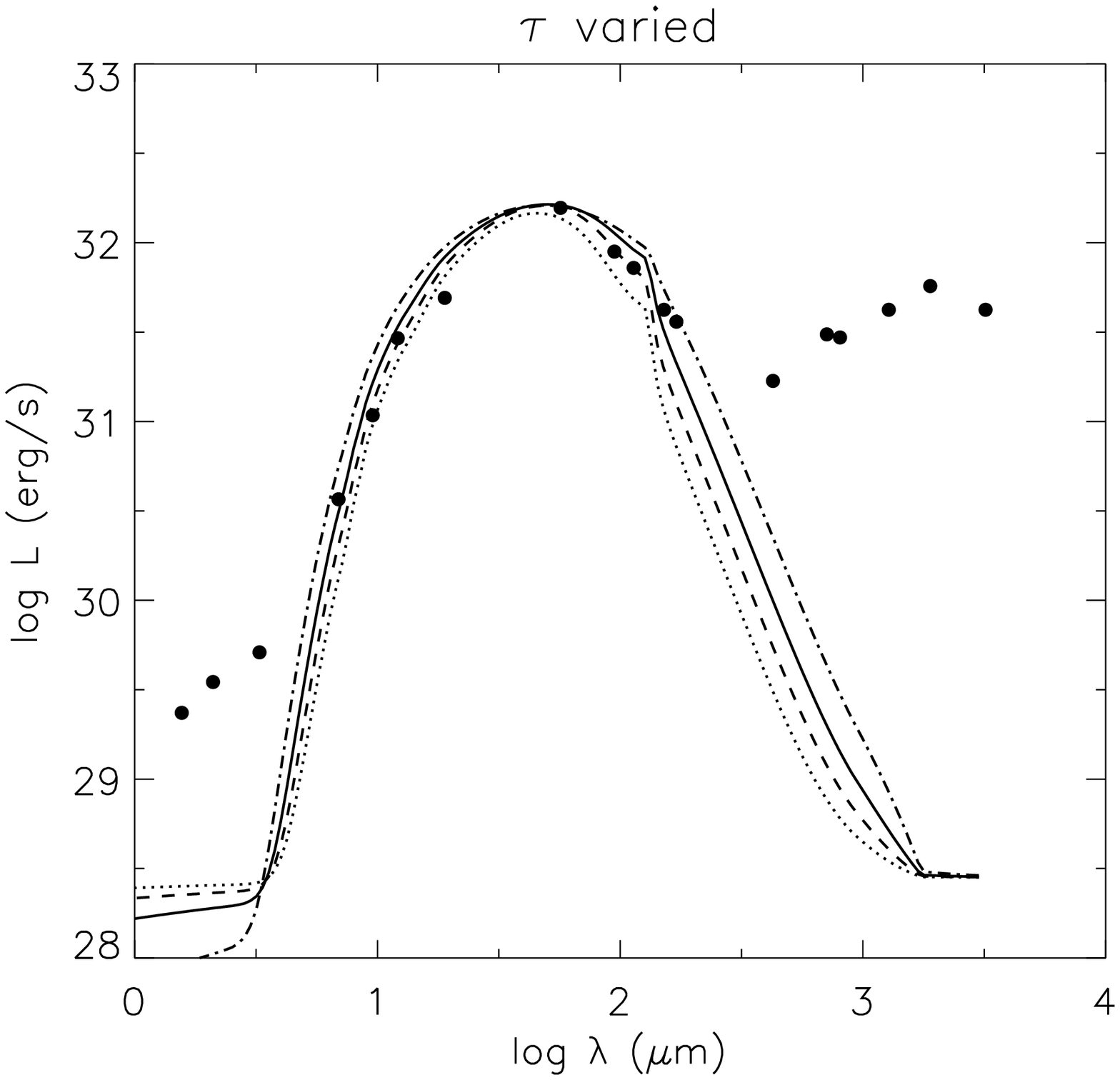}}
\resizebox{8cm}{!}{\includegraphics{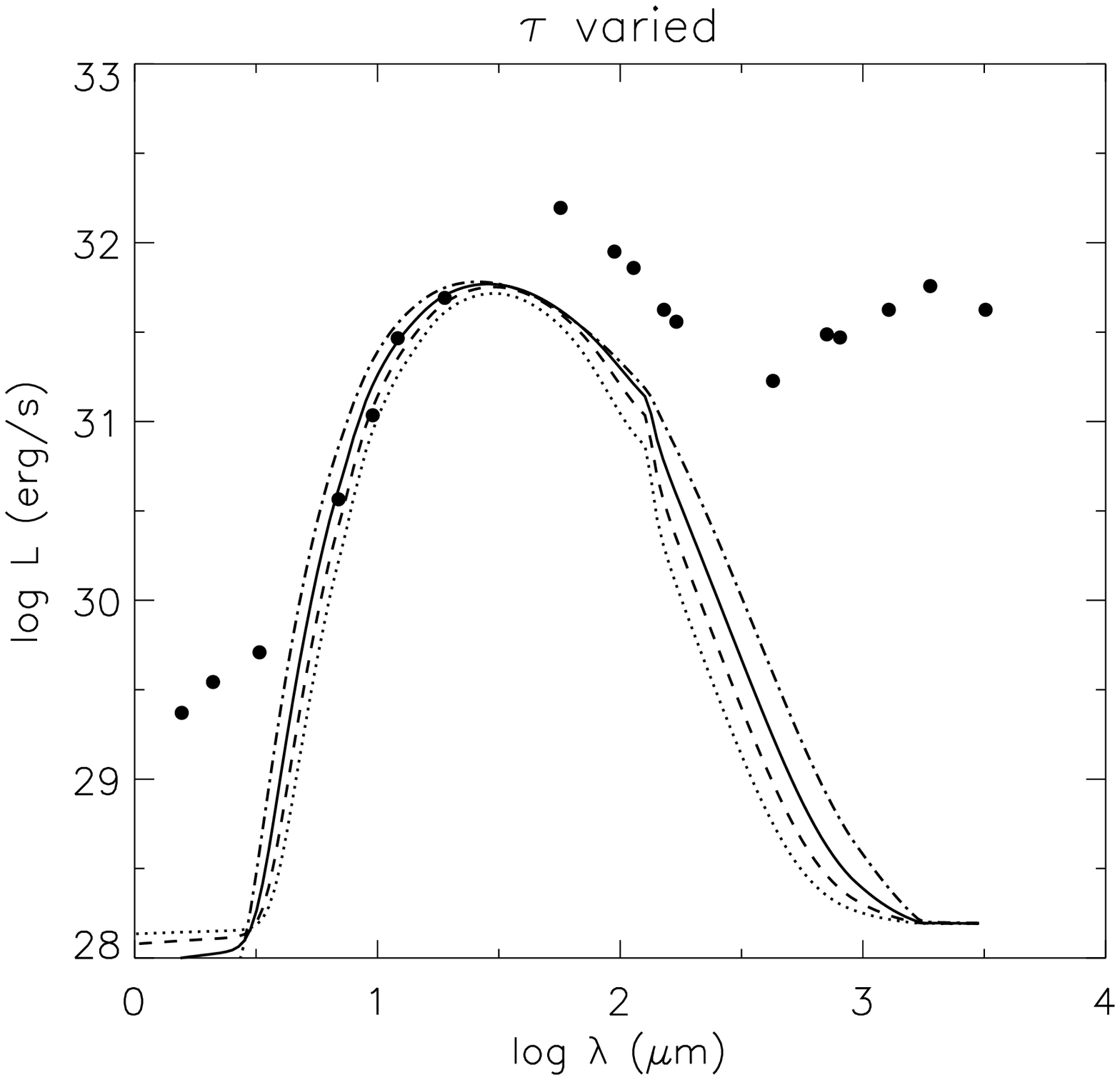}}

\resizebox{8cm}{!}{\includegraphics{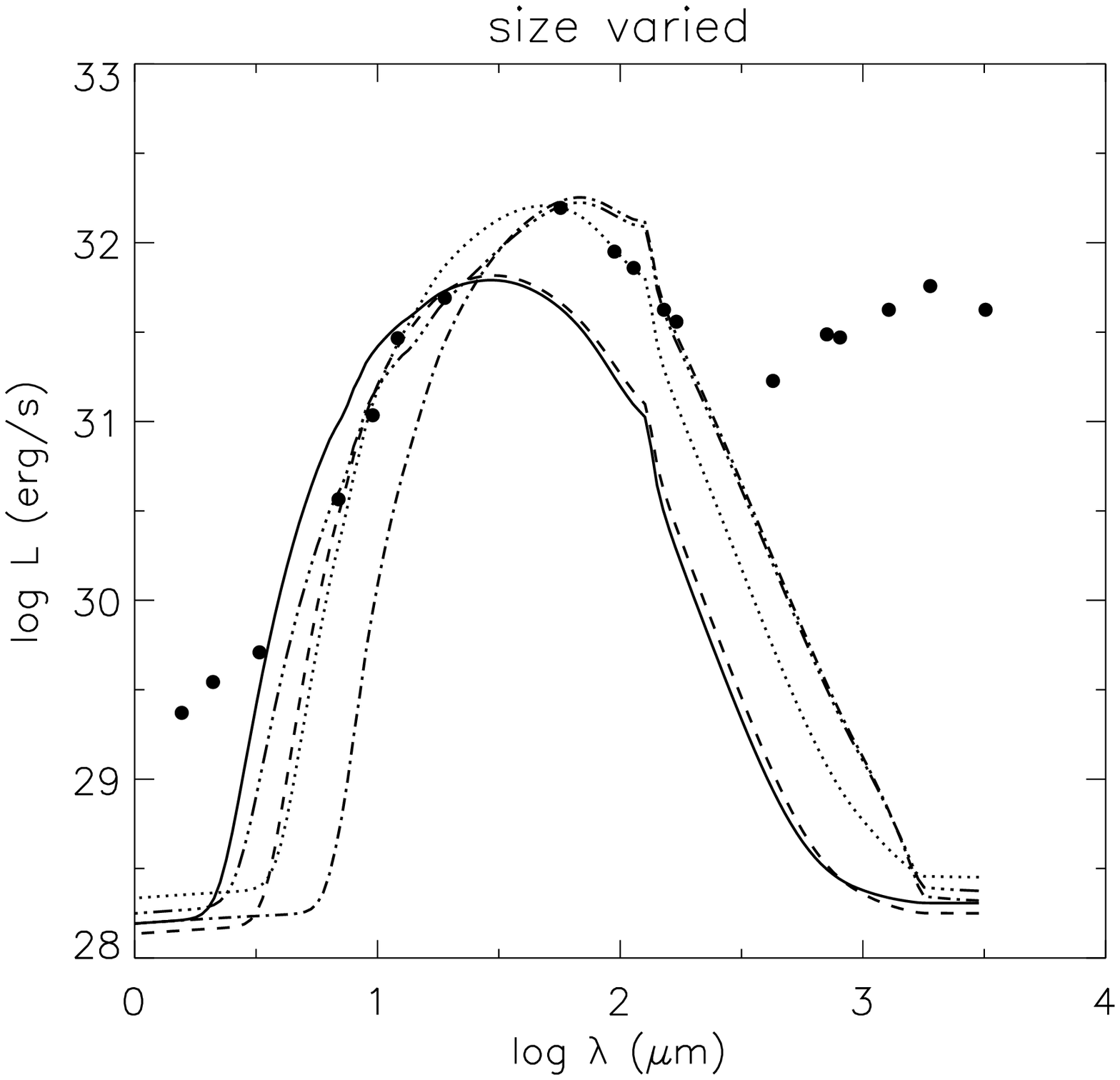}}
\resizebox{8cm}{!}{\includegraphics{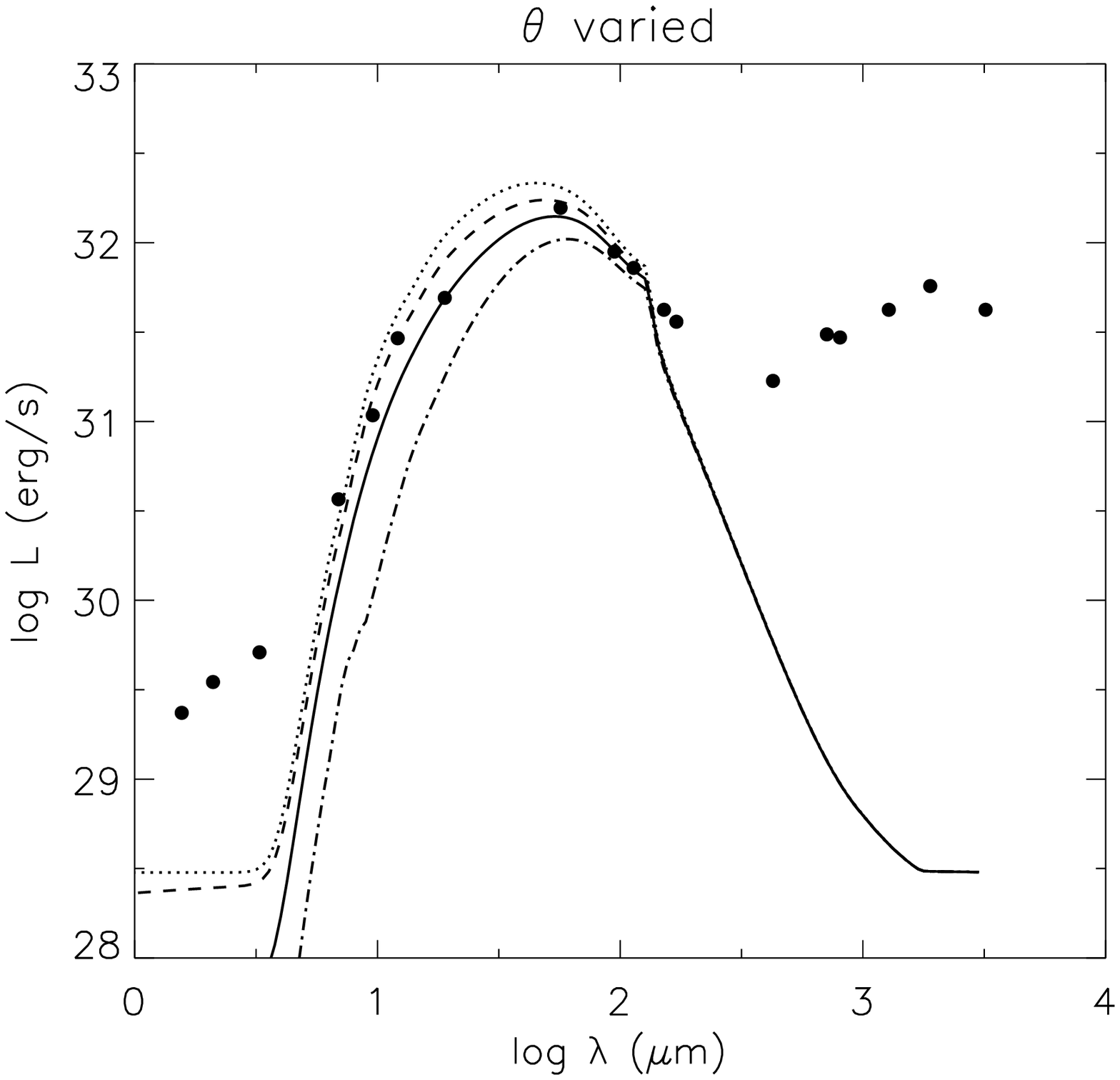}}
\caption{\label{7_fit_cl}
{\it Upper left panel}: model fits of a conical 
disk to the observed SED of \cyga. The error bars on the observed 
luminosities are smaller than the plot symbols. The model disk extends 
from 10 to 100\,pc from the nucleus and has an inclination of 
50$^{\circ}$. Lines are for different opacity models ($\tau_{20}$): 
5=dotted, 10=dashed, 20=solid and 50=dash-dotted. 
{\it Upper right panel}: as left panel, but for a disk from 10--30\,pc. 
{\it Lower left panel}: fits of disks with $\tau_{20}=10$ and inclination 
of 50 degrees. Lines are for different sizes: dotted=10--100\,pc, 
dashed=10--30\,pc, solid=3--30\,pc, dash-dot=30-300\,pc, dash-triple
dot=3-300\,pc.
{\it Lower right panel}: fits of disks with $\tau_{20}=10$, and size
10 to 100\,pc. Different lines are for different inclinations. From 
top to bottom: 30 (dotted), 50 (dashed), 60 (solid) and 70 (dash-dotted). 
All panels have $\alpha=1$ and $\beta=0$ models.
}
\end{center}
\end{figure*}

\subsubsection{Conclusions from colour-colour plots}
Most objects have negative cold spectral indices
which are not explained by the models. This could very well 
be due to a contribution from other components,
such as additional cold dust from star-formation regions.
Observational evidence for a second dust component heated 
by stars in active galaxies is found in several studies 
\citep{prie01,spi02,ivb01}. Therefore, we conclude that 
the torus emission only dominates
the SED to about 25\um, while the observed 60 and 100\um\ 
flux arise in a second infrared component. 

With respect to the unification theory,
there is no trend in the data for radio galaxies and Seyfert~2
to be more consistent with large inclination models, or
quasars and Seyfert~1 with small inclinations. In the hot vs warm plots 
all radio galaxies lie closer to the models than the quasars, while
the objects with known large star-formation rates occupy a clearly
distinct region, having cooler hot and warm colours.
The broad-line radio galaxy colours are consistent with the 
lack of large scale cold dust, as discussed in \citet{ivb01}. 
If the torus is the only infrared component in broad-line radio 
galaxies, they should also be consistent with the models
in their cold colours, but we lack detections at 100\um\
to confirm this. As emphasized
above, the differences between models and observations
are indicative of a second component in the far-infrared.

\subsection{Comparison with \cyga}
For individual objects, the model parameters can be constrained
by a direct fit to the observed infrared spectrum.
One of the best objects to test our models is the narrow-line
radio galaxy
\cyga, since it is the best observed radio-loud active galaxy in the
infrared, and has a rather well determined central continuum.
IRAS and ISO have observed \cyga\ in the infrared
and additional (sub-)millimeter data are available to
determine the non-thermal contribution. From optical observations
the size of the obscuring torus has been constrained to lie
between a few parsecs inner radius, up to maximum of a few hundred
parsecs outer radius \citep{ima00}. Also, we know that 
the angle of orientation of the radio jet is $\sim50$ degrees
\citep{car96}, allowing us to vary $\theta$
around this value. For the models we fix $\alpha=1$ and $\beta=0$,
since these are the best values from the colour-colour fitting.
We use $q=-2$ for the dust grain size distribution.

We have constructed models normalized to the observed 
hard X-ray luminosity and spectral slope, and taken into account
size limitations of the torus in \cyga. The inner and outer radius are 
varied with one extra value: $r_0=3$, 10 and 30\,pc, $r_{out}=30$,
100 and 300\,pc. Models are constructed for the same four values 
of $\tau_{20}$ and spectra are obtained for varying inclinations.\\

In Fig.~\ref{7_fit_cl} we show 
several fits to the spectrum of \cyga, with different
variables. A small scaling factor is
needed to fit the models, which is either due to a different
intrinsic luminosity, or different value of $h_0$, which are both 
kept fixed.

\subsubsection{Tuning the fit: size of the disk}
The size of the disk has the largest effect on the width of
the SED. Good fits are obtained for all optical depths for
a disk with $r=10-100$\,pc.
The reference model does not fit the SED, as expected from the
colour-colour analysis. It overproduces near-infrared emission, and
cannot account for all the far-infrared emission. The large and
outer disk can fit the far-infrared output of \cyga, but both
underpredict the observed mid-infrared flux. The inner
radius of the disk can be carefully tuned to the shape and
intensity of the mid-infrared observations. For an inner
radius of 10\,pc the mid-infrared is well fitted by
the model.

With an outer radius of 30\,pc
all four mid-infrared points are well fitted by the different
opacity models. The model with an outer radius of 100\,pc fits 
all the data, although the fit in the mid-infrared is somewhat worse.
The longest wavelength far-infrared point is only fitted by the 10--100\,pc
disk with $\tau_{20}=5$. Increasing the outer radius to 300\,pc 
improves the fit for other opacities at the longest wavelengths, 
but then the fit in the mid-infrared deteriorates.
The models with an outer radius of 100\,pc or more peak at 60\um, as 
observed, and the slope of the far-infrared and sub-millimeter tail 
is reasonably consistent with the observations.

We have no problem reproducing the width of the observed SED with a 
single large torus model, as did previous models. 
However, from optical imaging and spectropolarimetry 
we know that \cyga\ has large scale dust and star-formation 
\citep{jack98,fos99}, which is expected to 
contribute to the far-infrared emission. From the colour-colour
analysis we also conclude that the far-infrared emission beyond
60\um\ is generally not produced by a torus. Therefore, the 
10--30\,pc disk provides the best model for the torus in \cyga.

\subsubsection{Tuning inclination and opacity}
Small variations in inclination have little effect on the
resulting SED and models with inclinations between 5 and 60
degrees fit equally well. Models with higher inclinations
are ruled out.

The opacity cannot be fully constrained from the models. For the
10--100\,pc disk the $\tau_{20}=10$ and 20 models fit best,
which is also the case in the 10--30\,pc disk. However, in the
latter the opacity is not constrained by the far-infrared, which
leaves only four datapoints to optimize the fit.
The effect of varying $\alpha$ and $\beta$ has been described
above, and is similar in the \cyga\ models. Changing them will
not significantly improve the fit.

\subsubsection{Conclusions from the \cyga\ fit}
From the observations of \cyga\ we can constrain the size 
and inclination of
the obscuring torus. The geometry and density distribution are
fixed, varying these parameters would not significantly change
the output spectrum. Assuming significant star-formation in the
host galaxy produces the far-infrared emission, the best fit for
the torus is obtained
for a disk with size of 10--30\,pc, inclination of 50 degrees 
and optical depth of 10--20.
A single large disk fits all infrared points, but does not allow a 
contribution from the 
observed large scale dust, and is therefore not fully consistent 
with optical observations of young stars and large scale dust 
in \cyga. 

\section{Conclusions}
Radiative transfer models have been constructed to study the emission 
of a dusty torus around an active galactic nucleus. We have varied
optical depth, disk geometry, disk size, density distribution and
dust grain size distribution, to study the effect on the resulting 
SED. In most models, the dust becomes optically thin around
100\um, which implies that the emission is isotropic at longer
wavelengths. The models are compared to the observed infrared colours
of a sample of radio-loud active galaxies, and furthermore 
fitted to observations
of \cyga. The problem with previous models was often that the
resulting SED was not broad enough. We have identified this as a
matter of size and/or opacity: the larger the disk, or the higher
the opacity, the broader the temperature distribution, 
and, as a consequence, the broader the spectrum. In the fit
to \cyga\ we have no problem to reproduce the entire width of the
observed infrared SED.

From the comparison to observed colours, we conclude that the
tori in active galaxies are relatively large. The inner disk
model does not fit any of the observations: the inner radius
more likely lies around 10\,pc. The outer radius is not well
constrained. Pillbox models are rejected on basis of the
colour-colour analysis, both the conical and flaring disks 
are possible. The surface density seems constant with radius; for more
centrally concentrated models the colours are not consistent
with observations. Varying the optical depth can sometimes
improve the level of agreement, but its effect is diverse. 
An optimum value cannot be derived.

The objects that are best fitted by the models are the
radio galaxies. In the hot vs warm colours all broad-line 
radio galaxies fall entirely in the range of the model
colours. This is consistent with observational evidence that in
other objects a second dust component contributes, which
is not related to a central obscuring torus. This second
component alters the shape of the spectrum, and hence the 
colours, which explains the inconsistency of the models
with the observed cold colours. We conclude that the torus
dominates only at 12 and 25\um. As a consequence, selecting
active galaxies is best done using their 12 and/or 25\um\ 
power. However, at these wavelengths the quasars are
expected to be brighter, for which bias one has to correct.

\citet{hes95} find a relation between the 60\um\ luminosity
and radio power of active galaxies: the more powerful active
galaxies are more luminous at 60\um. They interpret this as
evidence that the 60\um\ emission is dominated by reprocessed
radiation from the obscuring torus. However, the above analysis 
suggests that the 60\um\ emission in active galaxies 
is dominated by a second component, and hence its power is not 
indicative of the power of the active nucleus. If star-formation 
dominates the 60\um\ emission in active galaxies, the correlation
would imply that the more powerful active galaxies have higher
star-formation rates. A better understanding of the process
dominating the 60\um\ emission can be obtained by comparing
individual fits to objects with and without star-formation, 
such as 3C\,48 and the broad-line radio galaxies.

For \cyga, the inner radius of the disk is constrained to
lie around 10\,pc. This value is found to be most consistent 
with the observations in the
colour-colour plots. The outer radius can be anywhere within
30--300\,pc, but the best fit is obtained with a value of
30--100\,pc. All values are within the observational limits set
by optical studies. Although a single torus model fits the entire 
infrared SED, there is strong observational evidence for a second
dust component in \cyga. In addition, the quality of the fit in 
the mid-infrared improves 
somewhat for a smaller disk. An outer radius of 30\,pc is preferred
to allow for the observed star-formation to contribute to the 
far-infrared emission in \cyga.

The consequences for unification
models are ambiguous. On one hand, we can fit the entire SED
of \cyga\ with a single torus model, while we know that other 
processes must contribute to its infrared emission. This is
in itself evidence in favour of the presence of an obscuring
torus. On the other hand, the cold colours of most objects 
are inconsistent with the torus models, indicating that a 
second component must be present, and a good fit to the 
entire SED is not possible. We know that this second component
must also be present in \cyga, and therefore, the relative
contribution of star-formation must be assessed with other
means to thoroughly disentangle the infrared SED of active
galaxies. Only with a proper decomposition of the SED can
the actual emission from an obscuring torus be studied in
detail.

\section{Future work}
Infrared observations provide a powerful tool to study the
emission from an obscuring torus in any type of active
galaxy, especially at near- and mid-infrared wavelengths, where 
the emission
from the torus is observed directly. The models presented in
this paper are a next step in increasing our understanding
of such tori. However, improvement of the models can be
expected on several fronts.

As mentioned, the 10\um\ absorption feature is sensitive to 
the parameters of the model. Unfortunately, no data presently
exist for radio-loud active galaxies, due to their faintness
at mid-infrared data and lack of capable instruments. SIRTF is
expected to have the sensitivity to observe a large sample
of radio-loud active galaxies, over a large range of redshifts.
By constructing models with 
higher frequency resolution around 10\um, predictions can be made
for future observations of 10\um\ emission or absorption
in radio-loud active galaxies. In addition the spectral index of 
the mid-infrared spectrum is sensitive to the parameters,
especially $\alpha_{12}^{25}$, and to some extent $\alpha_{25}^{60}$. 
Since this is the wavelength range where we are reasonably sure 
we observe emission from the torus, high resolution spectra
are needed. The SIRTF satellite will be able to obtain 
spectra up to 90\um\ for active galaxies over a large range
of redshifts, which will provide a good test for the models. 

The models presented here are not optimal. Improved models could
lead to a better understanding of the physics involved in the
obscuring torus around an active nucleus. They
could be made self-consistent with a scale height equal 
to the local radiative pressure scale height. The input spectrum could 
be adapted to better reflect the emission spectrum from an accretion
disk, which often shows a strong bump in the soft X-rays.
In addition, there is evidence that star-formation can occur 
in the torus, which will increase the scale height of the disk.
We plan to take this into account as well.
Finally, we have not treated indirect heating
by the hot gas around the accretion disk. This gas absorbs
the X-rays and re-emits the energy in the UV. This UV 
serves as an additional heating source for the dust.
However, improvement of the models is only relevant
when better observations are available to constrain
the models, and when we have a better understanding of the
contribution of star-formation to the infrared SED.

\acknowledgements

The authors wish to thank Xander Tielens for initiating the collaboration
and for advice and critical comments during the writing of the paper. 
Thanks to Peter Barthel for advice, comments and support, and thanks
to Julian Krolik for critical discussions. Finally, we acknowledge
expert reading by the referee, {\v Z}eljko Ivezi{\'c}, leading to 
significant improvement of this paper and many new ideas for the future.
CPD acknowledges support from 
the European Commission under TMR grant ERBFMRX-CT98-0195 (`Accretion 
onto black holes, compact objects and protostars'). 

\bibliographystyle{aa}
\bibliography{3761_bib}

\begin{thebibliography}{26}
\expandafter\ifx\csname natexlab\endcsname\relax\def\natexlab#1{#1}\fi

\bibitem[{{Canalizo} \& {Stockton}(2000)}]{can00}
{Canalizo}, G. \& {Stockton}, A. 2000, ApJ, 528, 201

\bibitem[{{Carilli} \& {Barthel}(1996)}]{car96}
{Carilli}, C.~L. \& {Barthel}, P.~D. 1996, A\&AR, 7, 1

\bibitem[{{Chiang} \& {Goldreich}(1997)}]{chia97}
{Chiang}, E.~I. \& {Goldreich}, P. 1997, ApJ, 490, 368

\bibitem[{{Dullemond} \& {Turolla}(2000)}]{dul00}
{Dullemond}, C.~P. \& {Turolla}, R. 2000, A\&A, 360, 1187

\bibitem[{{Efstathiou} \& {Rowan-Robinson}(1995)}]{efst95}
{Efstathiou}, A. \& {Rowan-Robinson}, M. 1995, MNRAS, 273, 649

\bibitem[{{Fosbury} {et~al.}(1999){Fosbury}, {Vernet}, {Villar-Mart{\'\i}n},
  {Cohen}, {Ogle}, \& {Tran}}]{fos99}
{Fosbury}, R.~A.~E., {Vernet}, J., {Villar-Mart{\'\i}n}, M., {et~al.} 1999, in
  The Most Distant Radio Galaxies, 311

\bibitem[{{Golombek} {et~al.}(1988){Golombek}, {Miley}, \&
  {Neugebauer}}]{gol88}
{Golombek}, D., {Miley}, G.~K., \& {Neugebauer}, G. 1988, AJ, 95, 26

\bibitem[{{Granato} \& {Danese}(1994)}]{gra94}
{Granato}, G.~L. \& {Danese}, L. 1994, MNRAS, 268, 235

\bibitem[{{Heckman} {et~al.}(1992){Heckman}, {Chambers}, \& {Postman}}]{heck92}
{Heckman}, T.~M., {Chambers}, K.~C., \& {Postman}, M. 1992, ApJ, 391, 39

\bibitem[{{Hes} {et~al.}(1995){Hes}, {Barthel}, \& {Hoekstra}}]{hes95}
{Hes}, R., {Barthel}, P.~D., \& {Hoekstra}, H. 1995, A\&A, 303, 8

\bibitem[{{Imanishi} \& {Ueno}(2000)}]{ima00}
{Imanishi}, M. \& {Ueno}, S. 2000, ApJ, 535, 626

\bibitem[{{Jackson} {et~al.}(1998){Jackson}, {Tadhunter}, \& {Sparks}}]{jack98}
{Jackson}, N., {Tadhunter}, C., \& {Sparks}, W.~B. 1998, MNRAS, 301, 131

\bibitem[{{Laor} \& {Draine}(1993)}]{laor93}
{Laor}, A. \& {Draine}, B.~T. 1993, ApJ, 402, 441

\bibitem[{{MacKenty}(1989)}]{mack89}
{MacKenty}, J.~W. 1989, ApJ, 343, 125

\bibitem[{{Maiolino} {et~al.}(2001{\natexlab{a}}){Maiolino}, {Marconi}, \&
  {Oliva}}]{maio01_2}
{Maiolino}, R., {Marconi}, A., \& {Oliva}, E. 2001{\natexlab{a}}, A\&A, 365, 37

\bibitem[{{Maiolino} {et~al.}(2001{\natexlab{b}}){Maiolino}, {Marconi},
  {Salvati}, {Risaliti}, {Severgnini}, {Oliva}, {La Franca}, \&
  {Vanzi}}]{maio01_1}
{Maiolino}, R., {Marconi}, A., {Salvati}, M., {et~al.} 2001{\natexlab{b}},
  A\&A, 365, 28

\bibitem[{{Mathis} {et~al.}(1977){Mathis}, {Rumpl}, \& {Nordsieck}}]{mrn77}
{Mathis}, J.~S., {Rumpl}, W., \& {Nordsieck}, K.~H. 1977, ApJ, 217, 425

\bibitem[{{Nenkova} {et~al.}(2002){Nenkova}, {Ivezi{\' c}}, \&
  {Elitzur}}]{nenk02}
{Nenkova}, M., {Ivezi{\' c}}, {\v Z}., \& {Elitzur}, M. 2002, ApJ, 570, L9

\bibitem[{{Neugebauer} {et~al.}(1986){Neugebauer}, {Miley}, {Soifer}, \&
  {Clegg}}]{neu86}
{Neugebauer}, G., {Miley}, G.~K., {Soifer}, B.~T., \& {Clegg}, P.~E. 1986, ApJ,
  308, 815

\bibitem[{{Pier} \& {Krolik}(1992)}]{pk92}
{Pier}, E.~A. \& {Krolik}, J.~H. 1992, ApJ, 401, 99

\bibitem[{{Prieto} {et~al.}(2001){Prieto}, {P{\' e}rez Garc{\' {\i}}a}, \&
  {Rodr{\' {\i}}guez Espinosa}}]{prie01}
{Prieto}, M.~A., {P{\' e}rez Garc{\' {\i}}a}, A.~M., \& {Rodr{\' {\i}}guez
  Espinosa}, J.~M. 2001, A\&A, 377, 60

\bibitem[{{Sekiguchi}(1987)}]{sek87}
{Sekiguchi}, K. 1987, ApJ, 316, 145

\bibitem[{{Spinoglio} {et~al.}(2002){Spinoglio}, {Andreani}, \&
  {Malkan}}]{spi02}
{Spinoglio}, L., {Andreani}, P., \& {Malkan}, M.~A. 2002, ApJ, 572, 105

\bibitem[{{van Bemmel} \& {Barthel}(2001)}]{ivb01}
{van Bemmel}, I.~M. \& {Barthel}, P.~D. 2001, A\&A, 379, L21

\bibitem[{{van Bemmel} {et~al.}(2000){van Bemmel}, {Barthel}, \& {de
  Graauw}}]{ivb00}
{van Bemmel}, I.~M., {Barthel}, P.~D., \& {de Graauw}, T. 2000, A\&A, 359, 523

\bibitem[{{van Bemmel} {et~al.}(1998){van Bemmel}, {Barthel}, \& {Yun}}]{ivb98}
{van Bemmel}, I.~M., {Barthel}, P.~D., \& {Yun}, M.~S. 1998, A\&A, 334, 799

\end{thebibliography}

\end{document}